\documentclass[aps,pra,twocolumn,superscriptaddress]{revtex4-1}
\usepackage{graphicx,bm,float,color,hyperref,amssymb,amsmath,physics,circuitikz,mymacros} %

\hypersetup{colorlinks=true linkcolor={blue}}

\begin{document}
\title{A tunable quantum dissipator for active resonator reset in circuit QED}
\author{Clement H. Wong}
\affiliation{Department of Physical Sciences, Arkansas Tech University, Russellville, AR 72801}
\author{Chris Wilen}
\affiliation{Department of Physics, University of Wisconsin-Madison, Madison, WI 53706}
\author{Robert McDermott}
\affiliation{Department of Physics, University of Wisconsin-Madison, Madison, WI 53706}
\author{Maxim G. Vavilov}
\affiliation{Department of Physics, University of Wisconsin-Madison, Madison, WI 53706}

\begin{abstract}
We propose a method for fast, deterministic resonator reset based on tunable dissipative modes.  The dissipator is based on a Josephson junction with relatively low quality factor.  When the dissipator is tuned into resonance with a high quality microwave resonator,  resonator photons are absorbed by the dissipator at a rate orders of magnitude faster than the resonator relaxation rate.  We determine the optimal parameters for realization of the tunable dissipator, and examine application of the dissipator to removing spurious photon population in the qubit readout resonator in circuit quantum electrodynamics. 
We show that even in the nonlinear large photon occupation regime, this enhanced resonator decay rate can be attained by appropriate modulation of the dissipator frequency.
\end{abstract}
\date{\today}
\maketitle

\section{Introduction}

Quantum error correction (QEC) demands fast, repetitive, and high fidelity measurement of ancilla qubits to detect errors \cite{*fowlerPRA12,*kellyNAT15,*ofekNAT16}. 
In circuit quantum electrodynamics (QED) systems, qubit measurement is performed by monitoring transmission of a microwave probe tone across a linear resonator that is dispersively coupled to the qubit {\cite{guPR17}}.
During the measurement, the resonator is populated with a large number of photons that must be removed before resuming gate operations; otherwise, the residual photons continue to measure and hence dephase the qubit \cite{*fowlerPRA12,*kellyNAT15,*ofekNAT16,gambettaPRA06}.
In many cases, resonator ring-down occupies a significant fraction of the QEC cycle.
Speeding up the resonator depletion rate is thus a challenging and important goal for QEC.  
Fast resonator reset is also important for quantum simulation {\cite{kandalaNAT17,georgescuRMP14}}.

Two approaches have been pursued for reducing the time needed to reset the resonator to its ground state.
Passive reset schemes use low quality $(Q)$ resonators together with a Purcell filter to inhibit qubit relaxation \cite{reedAPL10,jeffreyPRL14}; 
however, the depletion time, set by the photon leakage rate, is still limited by the size of the dispersive shift required for high fidelity readout \cite{yanNATC16}. 
Active reset methods use high $Q$ resonators and apply pulse sequences that remove photons from the resonator \cite{mcclurePRA16,bultinkPRA16}. 
However, complicated pulse sequences are necessary in the nonlinear regime, and depletion rates significantly faster than the bare resonator decay rate have yet to be achieved with this approach \cite{bultinkPRA16,boutinPRA17}.

In this paper, we propose a deterministic resonator reset scheme based on tunable dissipative modes.
The dissipator is formed by a strongly damped, frequency-tunable Josephson junction. 
When the dissipator is tuned to resonance with the readout resonator, the resonator relaxes at the rate of the dissipator mode, which can be significantly faster than previous proposals.  
When the dissipator is far detuned from the resonator, its damping of the readout resonator is minimal.  
We show that at the optimal the resonator-dissipator coupling, resonator photons can be depleted at {a fraction of} the dissipator damping rate.
Furthermore, we  show that  fast depletion of the resonator  can be attained in the nonlinear regime by appropriate modulation of the dissipator frequency. 
Our proposal has the advantage of simplicity, as it does not require sophisticated pulse sequences.
{As evidence of the utility of our proposed device, we note that in a recent experiment on qubit readout using photon counting, a method of photon depletion similar to that presented in this paper enabled repeated high-fidelity quantum nondemolition measurements \cite{opremcakSCI18}.  
We also note that a circuit dissipator formed with a low quality resonator has been studied in Ref.~\cite{partanenSR18}, and that a dissipator formed by Josephson junction chains was proposed in Ref.~\cite{rastelliPRB18}}

In addition to resonator reset, the circuit proposed here could be used to study driven-dissipative phase transitions of  coupled nonlinear oscillators \cite{carmichaelPRX15,*caoPRA16} and for quantum bath engineering \cite{murchPRL12,miranowiczPRA14}. It could also be relevant for engineering driven dissipative cat qubits {\cite{mirrahimiNJP14,*wangSCI16}}.

The central goal of this paper is to optimize the resonator photon number depletion rate as a function of the resonator-dissipator coupling $g$ for fixed dissipator relaxation rate $\g$.
That an optimal ratio of $g/\g$ exists can be seen by the following argument. 
When $g\simgt\g$, photons are exchanged between the resonator (mode $a$) and dissipator (mode $b$) via Rabi-like oscillations at the frequency $g$.  Since the photons spend half the time in the dissipator with damping rate $\g$, the depletion rate should be given by the average decay rate $\g/2$.
On the other hand, when $g\ll \g$, the dissipator mode is broadened into a continuum of states with linewidth $\g$,  so  that there will be a very small density of states resonant with resonator modes. 
{Here, relaxation occurs via the Purcell effect.}
The relaxation rate can be estimated from Fermi's golden rule, which gives the transition rate to the dissipator mode $\G_{a\to b}=(2\pi/\hbar)g^2\rho_b(\w_a)\simeq{g^2/\g}$, where $\rho_b(\w)\sim1/\g$ is the effective dissipator density of states  \cite{merzbacher60,*carmichael09} and $\w_a$ the resonator frequency. In this regime, the depletion rate decreases as a function of $\g$.  This argument suggests an optimal point at $\g\simeq g$. 
We will find that this estimate is quantitatively correct  in the linear regime.
For high photon numbers, the transition rate is suppressed as the resonant condition is shifted by  Kerr nonlinearity \cite{miranowiczPRA13}.  However, we will show this effect can be essentially eliminated with a compensating parametric pulse applied to the dissipator. 

\section{Model and Formalism}

\begin{figure}[t]
\includegraphics[width=\linewidth]{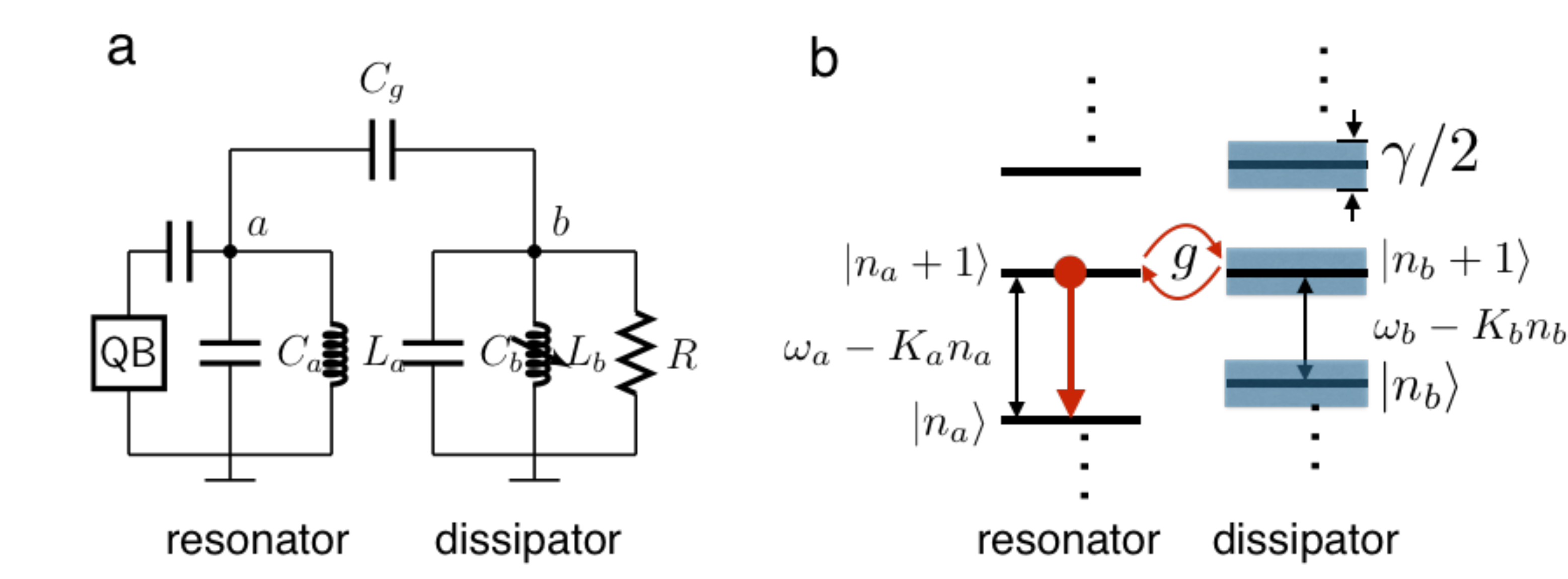}
\caption{(a) Circuit for deterministic resonator reset.  
Circuit $a$ on the left is the readout resonator coupled to a qubit.
Circuit $b$ on the right is the dissipator represented as a resonator with tunable inductance shunted by a resistor $R$. 
(b) Energy levels of the resonator (mode $a$) and the dissipator  (mode $b$).  The broadening of mode $b$ energy levels due to the large dissipator damping rate $\g$ is indicated by the blue shading.  Nonlinearity causes nonuniform energy spacings in both the resonator and the dissipator.}
\label{fig_circuit}
\end{figure}

For qubit readout, a microwave resonator is coupled to a qubit in the dispersive limit,  $\tilde g\ll\D_q$, where $\til g$ is qubit-resonator coupling and  ${\D_q=\w_q-\w_a}$ is the qubit-resonator frequency detuning.  In this limit, the qubit-resonator Hamiltonian projected onto the qubit logical states reads  \cite{mcclurePRA16,kochPRA07,*niggPRL12,*kirchmairNAT13,boissonneaultPRL10,*bishopPRL10}
\ben
{H_a'\over \hbar}=(\w_q-\chi a^\dag a){\s_z\over2}+\qty(\w_a-{\chi\over2})a^\dag a-{K_a\over2}a^\dag a^\dag aa~,
\label{Ha}
\een
where $\chi\approx-E_C/n_{\rm crit}$ is the qubit-induced dispersive shift of the resonator frequency, $K_a\simeq E_C/n_{\rm crit}^2$  is the qubit-induced self-Kerr coefficient of the resonator, and  $n_{\rm crit}=\D_q^2/4\tilde g^2$ is the critical photon number  .

We consider coupling the resonator to a dissipator made from a strongly damped ${Q_b\sim100}$, frequency-tunable Josephson junction.
The dissipator mode is weakly anharmonic and modeled by the Hamiltonian 
\ben {H_b\over\hbar}=\w_bb^\dag b-{K_b\over2}b^\dag b^\dag bb~,\een
where $b$ denotes  the destruction operator for the dissipator modes, $\w_b/2\pi$ is the tunable dissipator frequency, and  $K_b$ is  the dissipator Kerr coefficient.

The coupling between resonator and dissipator modes is governed by the Hamiltonian [see~Appendix \ref{app:circuit}]
\ben  {H_{ab}\over\hbar}=- g(a-a^\dag)(b-b^\dag)~.\label{Hab}\een
The system Hamiltonian is $H_{\rm sys}=H_a'+H_b+H_{ab}$.  The circuit diagram and its quantized modes are shown in \fig{fig_circuit}.

The coupling of the resonator to its input port and the dissipator to its external environment is modeled by the Hamiltonian
\begin{align}
{H_{\rm env}\over\hbar}&=\int {d\w\over2\pi} [\w (A^\dag_\w A_\w+B^\dag_\w B_\w)\nn
&-i\sqrt{\kap}(A^\dag_\w a-a^\dag A_\w)-i\sqrt{\g}(B^\dag_\w b-b^\dag B_\w)].
\label{bath}
\end{align}
Here, $\sqrt\kap$ is the input port coupling to an external transmission line used to drive the resonator,  we neglect internal losses in the resonator, and $A_\w$ and $B_\w$ are the transmission line and dissipator bath mode operators, respectively.
The dissipator's equivalent shunt resistance $R$ is represented as a coupling $\sqrt\g$ of the $b$ modes to the $B_\w$  modes of a semi-infinite transmission line with characteristic impedance $R$ \cite{voolIJCT17,yurkePRA84}.     
The dissipator environment is modeled by an equivalent shunt resistance {$R=1/\w_b C_b\tan\de$, where $\tan\de\simeq10^{-2}$ is the loss tangent}. The dissipator relaxation rate  $\g\gg\kap$ is related to the circuit parameters by $\g=1/RC_b$.  
The total Hamiltonian is $H=H_{\rm sys}+H_{\rm env}$.

\begin{figure}[t]
\includegraphics[width=0.7\linewidth]{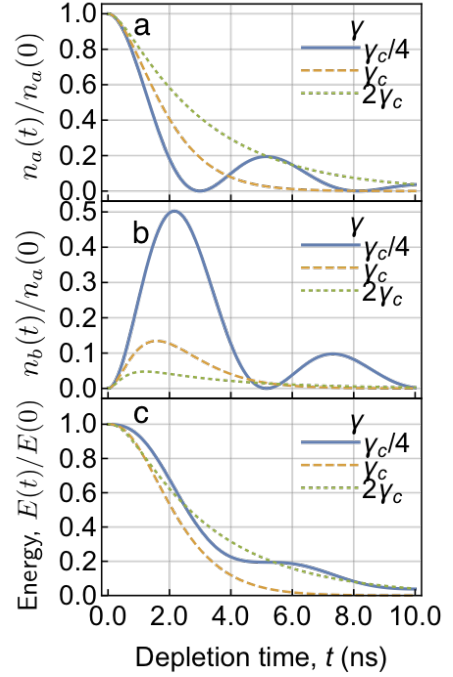}
\caption{Depletion of resonator and dissipator  photons in the linear regime.
Resonator (a) and  dissipator (b) photon number, and system energy (c) for the dissipator decay rates $\gamma=0.25\gamma_c$,  $\gamma_c$, and $2\gamma_c$, with  critical damping rate $\g_c=4g$ for a fixed coupling $g/2\pi=0.1$ GHz. }
\label{fig:decay}
\end{figure}

The Heisenberg-Langevin equations in the rotating wave approximation (RWA) are given by \cite{gardinerBook04}
\begin{align}
\dot{a}&=-\qty[{\kap\over2}+i(\w_a-K_a \hat n_a)]a-ig b-\sqrt{\kap}a_{\rm in}\label{EOM}\\
\dot{b}&=-\qty[{\g\over2}+i(\w_b-K_b\hat n_b)]b-iga-\sqrt{\gamma} b_{\rm in}\notag~,
\end{align}
where we have neglected small terms in the damping matrix of order $g/\w_a$; see Appendix \ref{app:eom}.
The dissipator $b$ is in equilibrium with a thermal bath $\ev{b^\dag_{\rm in}(\w) b_{\rm in}(\w')}\approx2\pi f_B(\w_b)\de(\w-\w')$, where $f_B(\w)=(e^{h\w/k_BT}-1)^{-1}$ is the Bose distribution at the temperature $T$. The bath modes thermalize at dilution refrigerator temperatures $T\sim10$ mK so that the dissipator initially has negligibly small occupation. The equations of motion \eq{EOM} contain all the physics we study in this paper.

\section{Linear dynamics of resonator  and dissipator modes\label{linear}}

\subsection{Mode damping}
We first consider the linear regime, setting $K_a=K_b=0$. Introducing the vector operators 
$\vec{X}=(a\,\,b)$ and $\vec{F}=(\sqrt{\kap}a_{\rm in}\,\, \sqrt{\g}b_{\rm in})$,  \eq{EOM} reads 
\ben\dot{\vec{X}}=-i(\bar\e+\vb M)\vec{X}-\vec{F}~,\label{Xdot}\een
where
\ben
\vb M=\begin{pmatrix}\D_b/2-i\g_-/4&-g\\-g&-\D_b/2+i\g_-/4
\end{pmatrix},
\een
${\Delta_b=\w_b-\w_a}$  is the detuning between the resonator and the dissipator, $\bar\e=\bar\w-i\bar\g/2$ is a complex parameter characterized by the average frequency
$\bar\w\equiv(\w_a+\w_b)/2$ and the average decay rate {$\bar\g\equiv(\kap+\g)/2$}, and $\g_-\equiv\g-\kap$  is the difference in decay rate. Below, we assume  $\bar\g,\g_-\simeq\g$.
The eigenvalues of the non-Hermitian matrix  $\vb M$ are given by \cite{milburnPRA15,rotterJPA09,PhysRevE.71.036227}
\ben \e_\pm=\bar\e\pm\D\e~,\quad\D\e=\sqrt{g^2+(z/2)^2}~,\label{eigval}\een
where $z=\D_b-i{\g_-/2}$.
The transformation that diagonalizes $\vb M$ is given by
\ben
\vb U=\begin{pmatrix} \cos(\eta/2)&-\sin(\eta/2)\\
\sin(\eta/2)&\cos(\eta/2)\label{U}\end{pmatrix},
\een
where $\tan\eta =g/z$, $\eta$ being a complex number \cite{milburnPRA15}.

\begin{figure}[t]
\includegraphics[width=\linewidth]{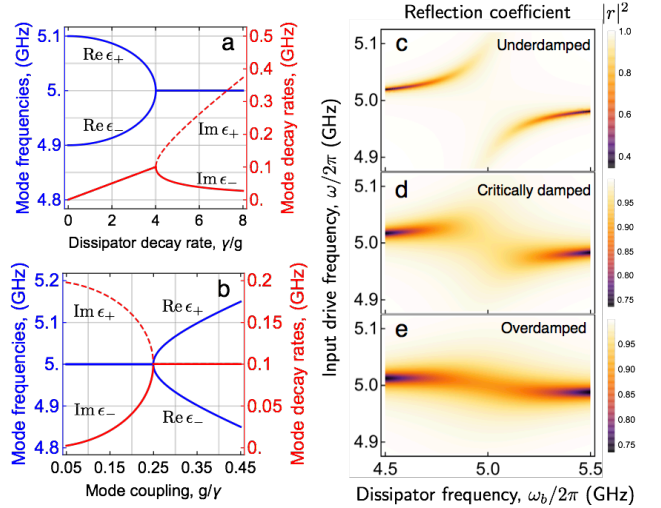}
\caption{Mode frequencies $\Re \e_\pm$ and decay rates $\Im \e_\pm$:
(a) as a function of dissipator decay rate normalized as $\g/g$ at a fixed coupling $g/2\pi=0.1$ GHz; 
(b) as a function of couping rate normalized as $g/\g$ at a dissipator decay rate $\g/2\pi=0.4$ ns$^{-1}$. 
In both cases, critical damping occurs at the cusp at $\g_c=4g$.
(c-e) Reflection coefficient of the resonator as a function of resonator drive frequency and dissipator frequency.}
\label{fig:eig}
\end{figure}

The solution to \eq{Xdot} is 
\ben\vec{X}(t)=\vb S(t)\vec{X}(0)+\de \vec{X}(t)~,\label{Xsol}\een
where we defined the response to input fields,
\ben\de \vec{X}\equiv-\int_0^t\,dt' \vb S(t-t')\vec{F}(t'),\een
{which is present even in the absence of drive due to input noise.}
The evolution operator 
\ben\vb S(t)\equiv e^{-i(\bar\e+ \vb M)t}= \vb U \begin{pmatrix}e^{-i\e_-t}  & 0 \\ 0 & e^{-i\e_+t} \end{pmatrix} \vb U^{-1}\label{S}\een
has matrix elements \footnote{Note that $S_{ij}\propto e^{-\bar\g t/2}$ has a damping envelope at the average decay rate.  In particular, when $\g_->g$ as $\D_b\to0$ and $g\to0$, the terms $e^{-\bar\g t/2}\sinh(\g_-t/4)$ and $e^{-\bar\g t/2}\cosh(\g_-t/4)$ remain finite   because $\bar\g>\g_-/2$.} 
\begin{align}
S_{aa}&=e^{-i\bar\e t}\qty[\cos (\D\e t)+i \cos \eta \sin (\D\e t)]\nn
S_{bb}&=e^{-i\bar\e t}\qty[\cos (\D\e t)-i \cos \eta \sin (\D\e t)]\nn
S_{ba}&=S_{ab}=ie^{-i\bar\e t}\sin \eta \sin (\D\e t),\label{Selem}
\end{align}
where
\ben
\cos\eta=\frac{z/2}{\sqrt{|z|^2/4+g^2}},\quad 
\sin\eta=\frac{g}{\sqrt{|z|^2/4+g^2}}.
\een
Equations \eqref{Xsol} and \eqref{Selem} completely determine the evolution of operators in the linear regime and  can be used to calculate all observables such as field amplitudes and correlations.
Substituting \eq{S} in \eq{Xsol} yields the  mode expansion
\ben
\vec{X}(t)=c_+e^{-i\e_+t} \vec{u}_++c_-e^{-i\e_-t}\vec{u}_-+\de \vec X~,\label{X}
\een
where $\vec u_-=(\cos(\eta/2),\sin(\eta/2))$ and $\vec u_+=(-\sin(\eta/2),\cos(\eta/2))$ are the eigenvectors of $\vb M$, and  we have defined the eigenoperators $(c_+,c_-)=\vb U^{-1}(a(0),b(0))$  \footnote{Note that, in contrast to normal modes, these dissipative modes are not orthorgonal because $\vb U$ is not a unitary matrix.}.  

We now consider photon relaxation for the case of a linear resonator. To remove photons from the resonator, we turn off input drives and tune the dissipator to the resonator frequency, so that  $\ev{\vec F(t)}=0$  and $\D_b=0$.
The mean occupation numbers are then  given by 
\begin{align}
&\bar{n}^{(0)}_i(t)=\expval{X^\dag_i(t)X_i(t)}\label{corr}\\
&=S^*_{ii'}(t)S_{ij'}(t)\expval{X^\dag_{i'}(0)X_{j'}(0)}+\ev{\de X^\dag_i(t)\de X_i(t)},\notag
\end{align} 
where $i=a,b$~, and  the supercript in $\bar{n}_i^{(0)}$ stands for zeroth order in Kerr coefficients $K_{a,b}$. 
The second term in \eq{corr} involves correlations due to quantum and thermal noise; this term is negligible compared to the first term.  
Furthermore, assuming negligible $b$ mode occupation and that the dissipator is initially decoupled from the resonator so that $a$ and $b$ modes are uncorrelated $\ev{b(0) a^\dagger(0)} =\ev{a(0) b^\dagger(0)}$, one finds 
\begin{align}
\bar{n}_a^{(0)}(t)&=|S_{aa}(t)|^2\bar{n}_a(0)~,\nn
\bar{n}_b^{(0)}(t)&=|S_{ba}(t)|^2\bar{n}_a(0)~.\label{eq:decay}
\end{align}

Photon number and energy relaxation for three decay regimes are plotted in Figs.~2a-c. The plots clearly show the underdamped, critically damped, and overdamped behavior according to \eq{eigval}, which gives the critical damping point $\g_c=\kap+4g$. 
In the underdamped regime $\g<\g_c$, the photon number undergoes damped oscillations with the average decay rate  $(\g+\kap)/2$ [cf.~\eq{underdamp}].
This regime can be explained physically by the fact that  photons spends an equal amount of the time in the resonator and the dissipator.   
The photons can thus be depleted essentially at the dissipator decay rate, as long as $\g\leq\g_c$.  When  $\g>\g_c$, the depletion rate decreases as a function of $\g$ because resonator photons are not efficiently transformed to dissipator excitations, as seen in $n_b(t)$ shown in \fig{fig:decay}b.
The maximum decay rate occurs at critical damping $\g=\g_c$, where there are no longer two distinct eigenvalues (see \eq{eigval}) and the matrix $\vb M$ is not diagonalizable.  
Such a degeneracy of the complex eigenspectrum of a non-Hermitian system is called an exceptional point \cite{*berryCJP2004,*dembowskiPRL01,*milburnPRA15}, analogous to an energy level crossing in Hermitian systems.
At this critical point, one finds from \eq{eq:decay}  (neglecting $\kap$)
\begin{align}
{\bar{n}^{(0)}_a(t)}&={\bar{n}_a(0)}e^{-2gt}(1+gt)^2~,\nn
{\bar{n}^{(0)}_b(t)}&={\bar{n}_a(0)}e^{-2g t}(gt)^2.\label{nacrit}
\end{align}
The resonator photon number $\bar{n}_a(t)$ can thus be depleted to  $0.1\%$  in $t_0\simeq 5.3/g$.
For $g/2\pi=0.1$  ns$^{-1}$, $t_0\simeq 8$ ns, which is significantly faster than currently achievable reset times \cite{jeffreyPRL14,bultinkPRA16}.

The real and imaginary parts of the eigenvalues in \eq{eigval}, corresponding to oscillation frequencies and damping rates of the photon field \eq{X}, are plotted in \fig{fig:eig} as a function of dissipator decay rate $\g$ at a fixed coupling $g/2\pi=0.1$ GHz for the case $\D_b=0$.
There are three different qualitative regimes similar to the underdamped, critically damped, and overdamped regimes of a single oscillator. 
At low damping $\g_-\ll4g$, there are two damped normal modes with a dissipation-dependent frequency splitting and a decay rate given by
\ben
\Re\e_\pm\simeq\bar\w\pm g\qty[1-\onehalf\pfrac{\g_-}{4g}^2]~,\quad\Im\e_\pm=\frac{\bar\g}{2}~.
\label{underdamp}
\een
At high damping, $\g_-\gg4g$, two decay modes emerge, with
\begin{align}
\Re\e_-&=\w_a~,\quad\Im\e_-\approx-\qty({\kap\over2}+{2g^2\over\g_-}),\nn 
\Re\e_+&=\w_b~,\quad\Im\e_+\approx-\qty({\g\over2}-{2g^2\over\g_-})~,\label{overdamp}
\end{align}
where $\e_-$($\e_+$)  governs the decay of the $a(b)$ mode. The dissipator-induced decay rate is $4g^2/\g$, consistent with the Fermi golden rule estimate made in the introduction.

Experimentally, one could tune the coupling at a fixed dissipator decay rate.  
One would then see the behavior in \fig{fig:eig}b, where the eigenvalues are plotted as function of $g/\g$ at $\g/2\pi=0.4$ ns$^{-1}$, for $\D_b=0$.  
The decay rate increases as a function of $g$ until it reaches the maximum in the underdamped regime given by average decay rate $\bar\g$.

We note that there are interesting adiabatic transport phenomena near the critical damping point in the $(g,\g)$ parameter space, which could enable transfer of photons from the resonator to the dissipator by adiabatically encircling this exceptional point \cite{uzdin11}.
While  adiabaticity can be achieved by staying sufficiently far away from the crossing point, the transport rate is expected to be slow.

\subsection{Resonator response to external drive}
Next, we analyze the response of the resonator to an external drive $\ev{a_{\rm in}}$. Generally, the response is determined by the retarded Green function
\ben\expval{a(t)}=-i\sqrt\kap\int dt'G_R(t-t')\ev{a_{\rm in}(t')},\een
or, in frequency space
\ben\expval{a(\w)}=-i\sqrt\kap G_R(\w)\ev{a_{\rm in}(\w)},\een
where $G_R(\w)$ generally has the form
\ben G_R(\w)={1\over \w-\w_a-i\kap/2-\Sigma(\w)}.\een
From the Fourier transform of \eq{EOM}, one finds $\Sigma(\w)=ig^2\chi_b$, where $\chi_b(\w)=i/(\w-\w_b+i\g/2)$ is the dissipator susceptibility \cite{clerkRMP10}. 
The mode frequencies and linewidth are then found from the poles $G_R^{-1}(\e)=0$, and they agree with \eq{eigval}.

The effective resonator damping rate can thus be tuned via the resonator-dissipator frequency detuning $\D_b$ according to \eq{eigval}.  
For example, in the overdamped regime, the resonator linewidth and frequency are given by [cf.~Appendix \ref{sec:aEOMeff}]
\begin{align} 
\kap'&=\kap+\g\frac{g^2}{\g^2/4+\D_b^2}~,\label{kapprime}\\
\w_a'&=\w_a-\D_b\frac{g^2}{\g^2/4+\D_b^2}~.\notag
\end{align}

Experimentally, the response function can be measured by applying a microwave drive and detecting the reflection amplitude as a function of input drive frequency $\w$,
\ben r(\w)=\frac{\ev{a_{\rm out}(\w)}}{\ev{a_{\rm in}(\w)}}={1-i\kap G_R(\w)}\label{r}\een
where we used the input-output relation $a_{\rm out}(\w)=\sqrt{\kap}a(\w)+a_{\rm in}(\w)$. The reflection coefficient $R=|r(\w)|^2$
is plotted in \fig{fig:eig}c-e as a function of input drive and dissipator frequency.  The resonances according to \eq{eigval} appear as dips in $|r(\w)|^2$.





\section{Photon depletion in the nonlinear regime}

To achieve fast and high-fidelity readout, it is necessary to use measurement photon numbers of order $n_{\rm crit}$, when nonlinear effects become important  \cite{jeffreyPRL14,bultinkPRA16}.  Optimal control techniques have been applied to
active reset in the presence of  this nonlinearity; however, this approach requires complicated pulses \cite{boutinPRA17}, and the depletion rates are still rather slow. 
On the other hand, as shown below, the method proposed here is applicable deep in the nonlinear regime, provided that mean field frequency shifts be compensated with simple exponential modulations  of the dissipator frequency.

Nonlinearity prevents the complete transfer of photons from the resonator to the dissipator. This effect can be understood in the mean field approximation, in which the interaction is linearized by approximating the occupation number operators in \eq{EOM} by their mean value $\hat n_a\simeq\bar{n}_a(t)=\expval{a^\dag(t)a(t)}$, $\hat n_b\simeq\bar{n}_b(t)=\expval{b^\dag(t)b(t)}$, leading to time-dependent oscillator frequencies
\ben\w_i(t)=\w_i-K_i\bar{n}_i(t),\een
where $i$ is the mode index.  The detuning due to this mean field shift reads
\ben\de\D_b(t)=K_a\bar{n}_a(t)-K_b\bar{n}_b(t).\een
The nonlinear energy levels are illustrated in \fig{fig_circuit}b.  
Typically, nonlinear effects become important around $n_{\rm crit}$. 
However, the resonant condition is relaxed by the large dissipator linewidth, so that nonlinear effects become important only when $\de\D_b\simeq\g$. 
For $K_a\sim1$ MHz and $\g/2\pi=0.1$ ns$^{-1}$, this yields $n_a\sim100$, which can be significantly larger than $n_{\rm crit}$. 

In the mean field approximation, the mean occupation numbers $\bar{n}_{a,b}$ can be solved  self-consistently in a perturbative approach. The solutions $\bar{n}_i^{(1)}(t)$ to  first order in $K_{a,b}$, are found by solving \eq{EOM} using the zeroth order solution $\bar{n}^{(0)}_{a,b}(t)$ given by  \eq{eq:decay}, and this process can be iterated to the desired accuracy. This suggests that  one can speed up the photon decay rate by modulating the oscillator frequencies as 
\[\w_i(t)=\w_i+K_i\bar{n}^{(0)}_i(t),\]
which compensates for the Kerr frequency shifts, maintaining the resonant condition.  

\begin{figure}[t]
\includegraphics[width=\linewidth]{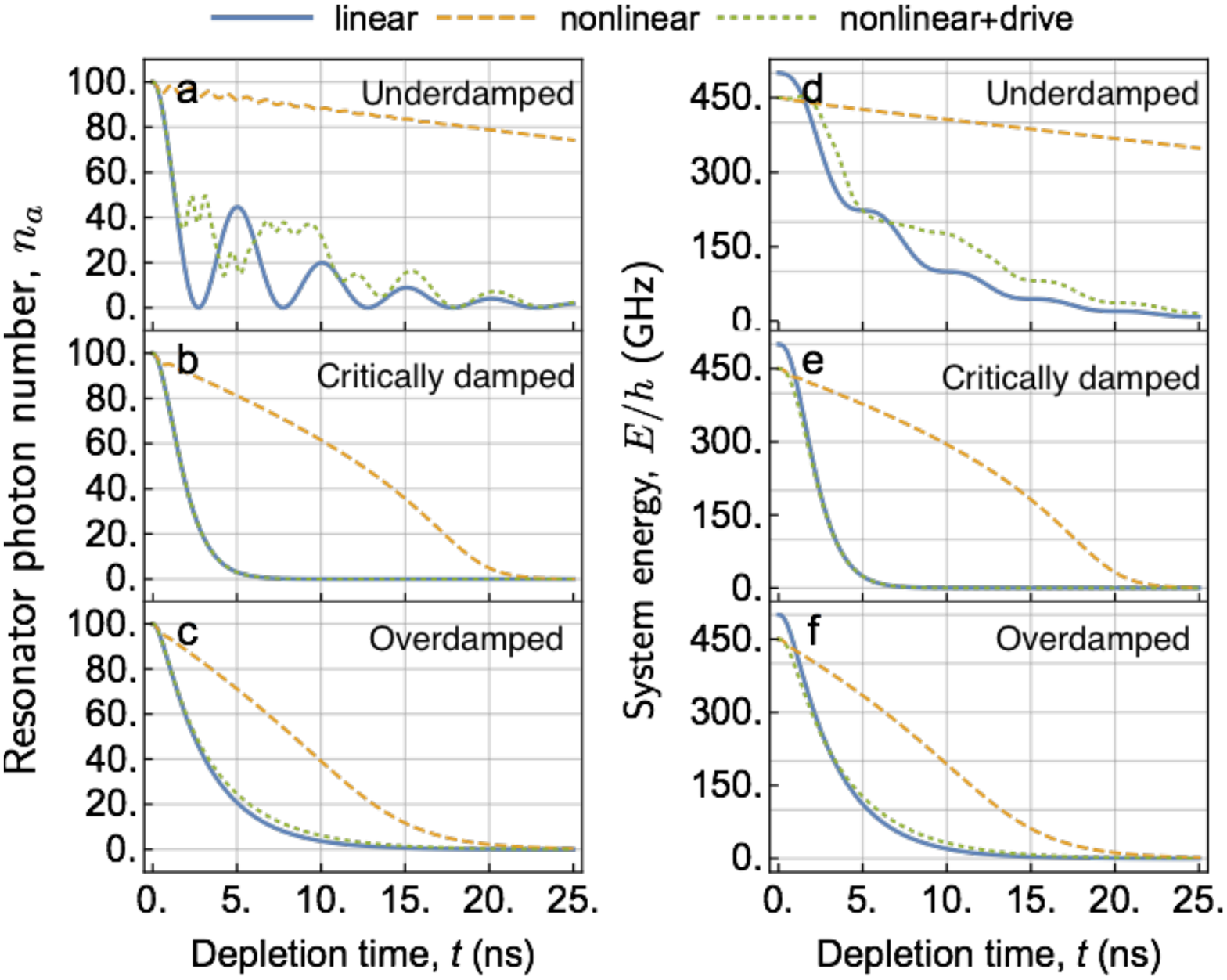}
\caption{Photon depletion in the nonlinear regime:  Resonator photon number (a-c) and total energy  (d-f) [cf.~\eq{E}] of the resonator-dissipator system as a function of time, computed from numerical solution of the semiclassical equations \eq{semiclass} for $g/2\pi=0.1$ GHz, $K_a/2\pi=10$ MHz, $K_b/2\pi=25$ MHz, and the dissipator decay rates in the three damping regimes,
 $\g/2\pi=(0.05,0.4,0.8)$ ns$^{-1}$.
The initial average resonator photon number is $\bar{n}_a(0)=100$. 
 In each plot, the linear solution is shown as blue solid lines.  
 The nonlinear solutions  with  and without optimization by the applied pulse \eq{pulse} on the dissipator frequency are plotted as  orange dashed lines and green dot dashed lines, respectively.}
\label{fig:scdecay}
\end{figure}     

We are thus led to the following optimization procedure. Since in practice the resonator frequency is fixed while the dissipator frequency is tunable, we apply a pulse on the dissipator frequency $\w_b(t)=\w_b^0+\de\w_b(t)$, where 
\begin{align}
\de\w_b(t)&\equiv K_b\bar{n}^{(0)}_b(t)-K_a\bar{n}^{(0)}_a(t)\nn
&=(K_b|S_{ba}(t)|^2-K_a|S_{aa}(t)|^2)\bar{n}_a(0)\label{dwb}.
\end{align}
Here, $S_{ij}$ are given in \eq{dwb}.  We find it sufficient to simplify these applied pulses as follows for the underdamped, critically damped, and overdamped case: 
\begin{align}
&\de\w^{({\rm under})}_b(t)=\bar{n}_a(0)e^{-\gamma t/2}(K_b\sin^2 gt-K_a\cos^2 gt),\nn
&\de\w^{({\rm crit})}_b(t)=\bar{n}_a(0)e^{-2gt}[K_b(gt)^2-K_a(1+gt)^2],\nn
&\de\w^{({\rm over})}_b(t)=\bar{n}_a(0)\nn
&\quad\times\qty[K_b\pfrac{4g}{\g}^2e^{-\g t/2}\sinh^2(\g t/4)-K_ae^{-4g^2t/\g}]~.\label{pulse}
\end{align}



\begin{figure}[t]
\includegraphics[width=0.8\linewidth]{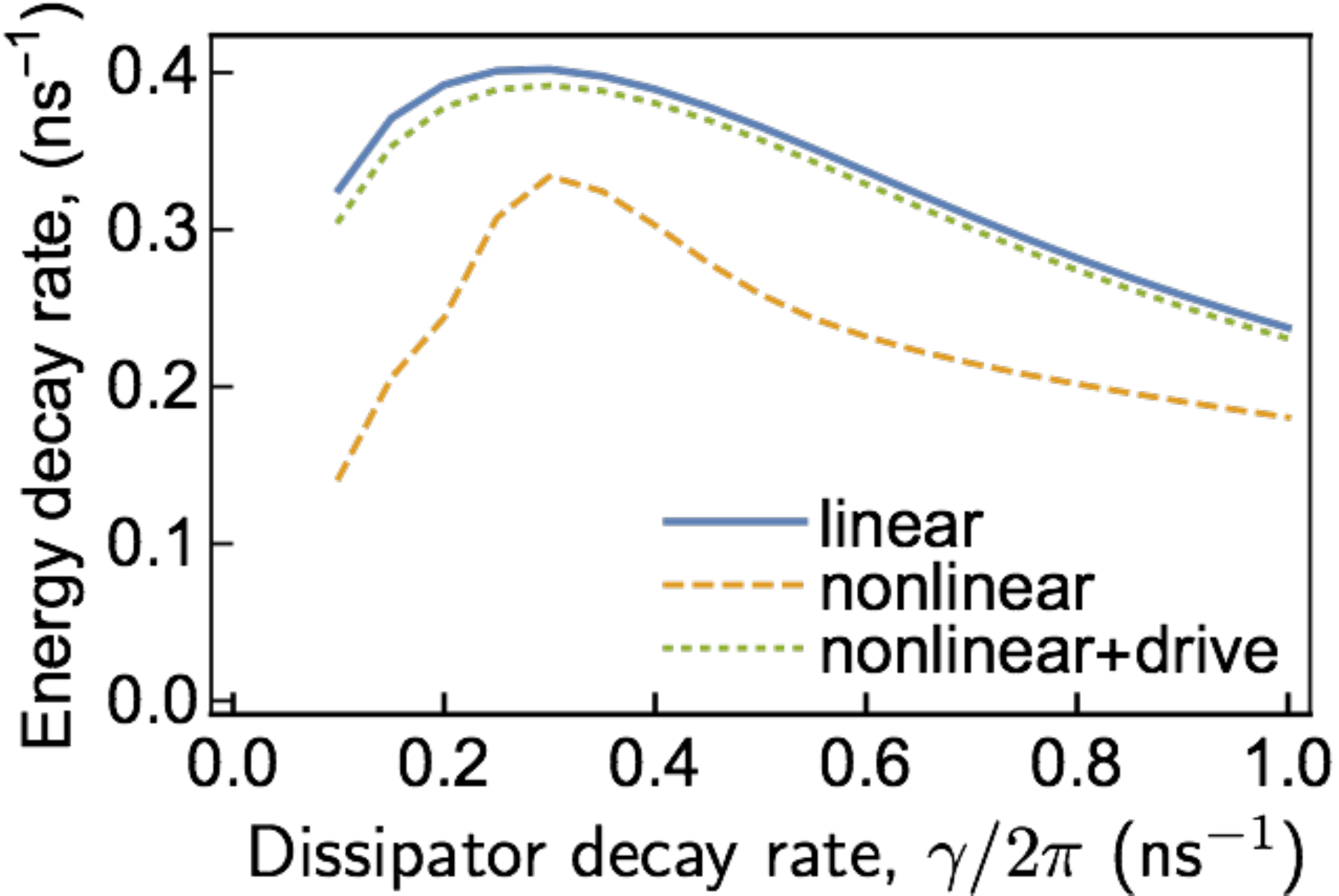}
\caption{Energy relaxation rate $1/T_1$ plotted as a function of the dissipator decay rate $\g/2\pi$ for the linear, nonlinear, and optimized nonlinear resonator reset.  Here, we took $g/2\pi=0.1$ GHz, $K_a/2\pi=5$ MHz, $K_b/2\pi=25$ MHz.}
\label{fig:T1}
\end{figure} 

\subsection{Semiclassical equations of motion}

First, we consider the semiclassical approximation, which is valid at the large photon numbers where  nonlinear effects become important. Taking the expectation value of \eq{EOM} and factorizing all correlators yields the semiclassical equations
\begin{align}
\dot{\al}&=-\qty[{\kap\over2}+i\qty(\w_a-K_a|\al|^2)]\al-ig\be\nn
\dot{\be}&=-\qty[{\g\over2}+i\qty(\w_b-K_b|\be|^2)]\be-ig\al~,\label{semiclass}
\end{align} 
where $\al\equiv\ev{a}$ and $\be\equiv\ev{b}$. The total system energy is
\ben
{E\over \hbar}=\w_a|\al|^2+\w_b|\be|^2+ g(\al\be^*+\be^*\al)-{K_a\over2}|\al|^4-{ K_b\over2}|\be|^4~.
\label{E}
\een
The photon number and total energy decay computed from \eq{semiclass} are plotted in \fig{fig:scdecay}(a-f), including the optimization pulse \eq{pulse}, {for the  parameter values $K_a/2\pi=10$ MHz and $K_b/2\pi=25$ MHz. }

In the underdamped regime, resonator photons are ``self-trapped" and cannot reach the dissipator \cite{wallsBook2012,eilbeckPD85}.
In the critically and overdamped regimes,  there is still a significant slow-down of the mode depletion.
In all cases, however, the slow-down due to nonlinearity can be essentially eliminated using the optimization pulse.

These results are summarized in \fig{fig:T1}, where the energy relaxation rate $1/T_1$, defined by the first time the energy decreases by the factor $E(T_1)/E(0)=e^{-1}$, is plotted as a function of the dissipator decay rate $\g/2\pi$.  
Here, we take $K_a/2\pi=5$ MHz, and the optimized energy relaxation is computed with the exact expressions in \eq{dwb}. 
The relaxation time is shortest in the underdamped regime because resonator photons can be completely transferred to the dissipator after a half period $\pi/g$, the maximum physical transfer rate.
However, to prevent photons from returning to the resonator would require perfect timing in switching off the dissipator. 
\begin{figure}[t]
\includegraphics[width=\linewidth]{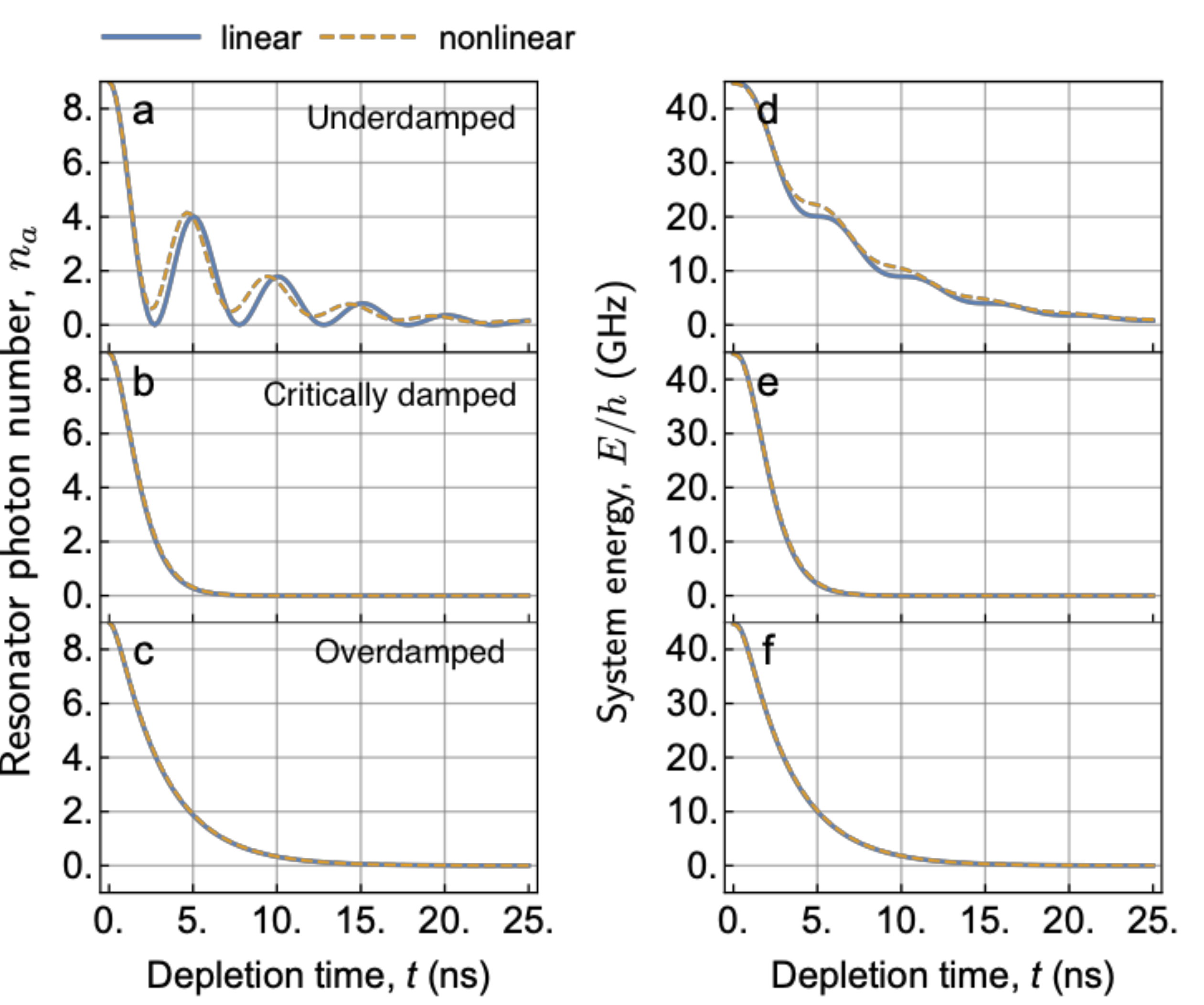}
\caption{Photon depletion in the nonlinear regime:  Resonator photon number (a-c) and total energy  (d-f) \eq{E} of the resonator-dissipator system as a function of time, computed from numerical solution of  the density matrix master equation \eq{master}, for $g/2\pi=0.1$ GHz, $K_a/2\pi=10$ MHz, $K_b/2\pi=25$ MHz, and the dissipator decay rates in the three damping regimes, $\g/2\pi=(0.05,0.4,0.8)$ ns$^{-1}$.
The initial average resonator photon number is $\bar{n}_a(0)=9$. 
 In each plot, the linear solution is shown in blue solid line.  The nonlinear solution is plotted as an yellow dashed line.}
\label{fig:qudecay}
\end{figure}     

\subsection{Quantum master  equations}

To study photon relaxation for small photon numbers, where quantum effects may be important, we use the density matrix master equation (cf.~appendix \ref{app:master})\footnote{Note that even with $\sim10$ initial photons, the basis required for this simulation has $\sim10^2$ states, and \eq{master} involves $\sim10^4$ coupled differential equations.}
\ben\frac{\p\rho}{\p t}=-\frac{i}{\hbar}[H_{\rm sys},\rho]+[\kap\mcal{D}(a)+\g\mcal{D}(b)]\rho\equiv\mcal L[\rho]\,,\label{master}\een
where $\mcal{D}(x)\rho=(2x\rho x^\dag-x^\dag x\rho-\rho x^\dag x)/2$ is the Linblad superoperator, and $\mcal L$ is the Liouvillan superoperator.
The solution  can be formally expressed as $\rho(t)=e^{\mcal L t}\rho(0)$, and the nonlinear dissipative mode frequency and decay rates are now defined by eigenvalues of $\mcal L$.  However, $\mcal L$ has $(n_an_b)^4$ matrix elements which makes diagonalization a difficult numerical task.
Instead, we solve for the decay dynamics of \eq{master} numerically in the uncoupled basis $\{\ket{n_an_b}\}$ using  QuTiP \cite{johanssonCPC12}.  
The results are plotted in \fig{fig:qudecay} (a-f), along with the plot of the linear solution $K_a=K_b=0$, {for the  parameter values $K_a/2\pi=10$ MHz and $K_b/2\pi=25$ MHz. }
The decay behaviors of the linear and nonlinear case are essentially identical because at these small photon numbers, nonlinear effects are negligible.  
Furthermore, \fig{fig:qudecay} (b) confirms that fast reset of approximately 8 ns can be achieved, consistent with the analytic solution to the linear equations of motion in section \ref{linear}.

\section{Conclusion and discussion}

We have proposed a resonator circuit with tunable dissipation derived from coupling to a damped Josephson mode.
In the optimal parameter regime, we show that resonator photon depletion times orders of magnitudes shorter than the intrinsic  resonator decay time are achievable.  
In the nonlinear regime, we have shown that this depletion time scale persists, as long as we compensate for mean field frequency shifts  by modulating the dissipator frequency.


A related problem to the one studied here is qubit initialization.  
This could be done by simply setting the qubit on resonance with the resonator in its low $Q$ state \cite{tuorilaNPJ17}.
Alternatively, one could  initialize the qubit into its ground state by driving the red sideband transition from $\ket{e,0}\to\ket{g,1}$, which converts the qubit excited state into a photon that  then decays quickly at the rate $\kap'$ \cite{blaisPRA07,wallraffPRL07,magnardPRL18}.  This problem deserves a separate study.

\section{ACKNOWLEDGMENTS}

We are thankful to {Naveen Nehra}, Konstantin Nesterov and {Alex Opremcak}  for fruitful discussions.  This work at the University of Wisconsin-Madison was supported by the U.S. Government under ARO Grants W911NF-14-1- 0080 and W911NF-15-1-0248.

\appendix

\section{Derivation of the linear system Hamiltonian \label{app:circuit}}

We model the resonator (circuit $a$) coupled to dissipator (circuit $b$) as two $LC$ circuits coupled by a capacitor $C_g$, as shown in \fig{fig_circuit}a, 
\begin{align}
\mcal L=\onehalf \dot{\vec\Phi}\hat C\dot{\vec\Phi}-\onehalf \vec \Phi \hat L^{-1} \vec \Phi~,\label{lagrange}
\end{align}
where we neglect nonlinearities.
Here, $\vec \Phi=(\Phi_a,\Phi_b)$, and the capacitance and inductance matrices are defined as
\ben
\hat C=\begin{pmatrix}C_a+C_g&-C_g\\-C_g&C_b+C_g\end{pmatrix},~~ 
\hat L^{-1}=\begin{pmatrix}L_a^{-1}&0\\0&L_b^{-1}\end{pmatrix}~.\label{CLmatrix}
\een
The charges on each node are given by cannonical momenta $Q_i={\p \mcal L/\p\dot \Phi_i}= C_{ij}\dot\Phi_j$. Legendre transformation yields the Hamiltonian
\begin{align}
H&=\vec{Q}\cdot\dot{\vec\Phi}-\mcal L=\onehalf \vec Q \hat C^{-1} \vec Q+\onehalf \vec \Phi \hat L^{-1} \vec \Phi\nn
&=\sum_{i=a,b}{\w_i^2L_iQ_i^2\over2}+{\Phi_i^2\over2L_i}+\be\sqrt{L_aL_b}\w_a\w_bQ_aQ_b~,
\label{H}
\end{align}
where  
\begin{align}
\w_i&=\sqrt{C^{-1}_{ii}/L_{i}}=\w_i^{(0)}\sqrt{C^{-1}_{ii}C_i}\label{omega}\\
C^{-1}_{aa}C_a&={1+C_g/C_b\over 1+C_g(C_a^{-1}+C_b^{-1})}\nn
C^{-1}_{bb}C_b&={1+C_g/C_a\over 1+C_g(C_a^{-1}+C_b^{-1})}\notag
\end{align}
are the mode frequencies including renormalization by the coupling $C_g$, $\w_i^{(0)}=1/\sqrt{L_iC_i}$ are the uncoupled $LC$ resonant frequencies,
and we have defined 
\[\beta=\frac{C_{ab}^{-1}}{\sqrt{C_{aa}^{-1}{C_{bb}^{-1}}}}=\frac{C_g}{\sqrt{(C_a+C_g)(C_b+C_g)}}.\]
The Hamiltonian equations reads
\begin{subequations}\begin{align}
\dot \Phi_i&={\p H\over \p Q_i}= C^{-1}_{ij}Q_j\label{Peom}\\
\dot Q_i&=-{\p H\over\p \Phi_i}=-{\Phi_i\over L_i}.\label{Qeom}
\end{align}\end{subequations}

\begin{figure}[t]
\includegraphics[width=\linewidth]{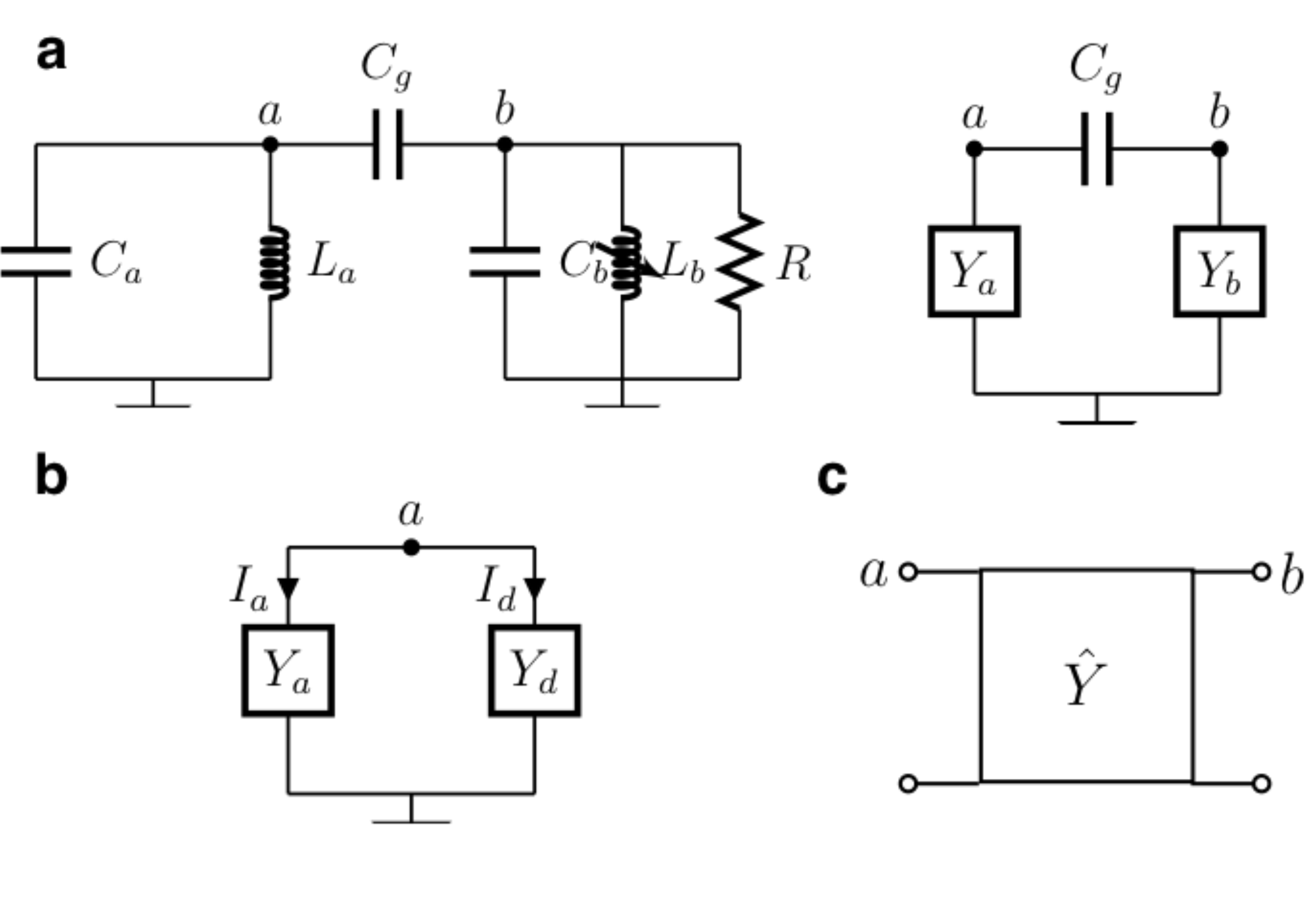}
\caption{(a) (left) Circuit diagram for the resonator-dissipator system  (right) Circuit diagram defining the resonator and dissipator admittances.
(b)  Equivalent circuit for the resonator in parallel with an admittance $Y_d$, see \eq{Yd1}.
(c)  Equivalent circuit for two port admittance matrix $\hat Y$, see \eq{Yhat}.}
\label{fig:admit}
\end{figure}

Quantizing the circuit, the flux and charge operators obey $[Q_i,\Phi_j]=-i\hbar\de_{ij}$. These operators are expressed in terms of the mode operators as follows:
\begin{align}
\Phi_a&=\sqrt{\hbar Z_a\over2}(a+a^\dag)\nn
Q_a&=-i\sqrt{\hbar\over2Z_a}(a-a^\dag)~,
\label{eq:modes}
\end{align}
where $Z_a\equiv\w_aL_a=\sqrt{L_a C^{-1}_{aa}}$, $[a,a^\dag]=1$, and similarly for the $b$ modes. The Hamiltonian then becomes
\begin{align}
{H\over\hbar}&=\w_a(a^\dag a+\onehalf)+\w_b(b^\dag b+\onehalf)\nn
&- g(a-a^\dag)(b-b^\dag)~,\label{eq:Hg1}
\end{align}
where the coupling is 
\begin{subequations}\begin{align}
g={\beta\sqrt{\w_a\w_b}\over2}
&=\frac{C_g\sqrt{\w_a\w_b}}{2\sqrt{(C_g+C_a)(C_g+C_b)}}.
\label{app:g1}
\end{align}\end{subequations}
 In the limit ${C_g\ll C_{a,b}}$, we find to leading order in $C_g$
\begin{subequations}\begin{align}
\w_i&\approx \omega_i^{(0)}\qty(1-\frac{C_g}{2 C_i})~,\quad i=a,b \label{app:freq}\\
g&\approx{C_g\over2}{\sqrt{\omega _a\omega _b\over C_aC_b}}\label{app:g2}
\end{align}\end{subequations}
Taking $C_a=C_b$, $\w_a^{(0)}=\w_b^{(0)}=\w_0$, we have
\ben\frac{C_g}{C_b}\approx\frac{2 g}{\omega _0}+\pfrac{2 g}{\omega _0}^2.\een
Below, we will use $\order{C_g/C_i}$ and $\order{g/\w_0}$ interchangeably.
We also note here that $g$ has a finite limit as $C_g\to\infty$, which yields  $\beta\to1$ and
\ben g_{\rm max}=\frac{1}{2 \sqrt{(C_a+C_b)\sqrt{L_a L_b}}}\label{eq:g2}\een
For $C_a=C_b$ and $L_a=L_b$, $g_{\rm max}=\w_0/2\sqrt2$.

\section{Circuit equations of motion\label{app:eom}}

In the main text, we modeled our system with equations of motion in the RWA and kept dissipator damping terms to leading order in $g/\w_0$.  
In this appendix, we present the linear circuit equations of motion without approximations.  
We show that they lead to the dissipative mode spectrum discussed in Appendix \ref{app:eom}, and reduce to the spectrum in the main text in the appropriate limits.
We compute numerically the  dynamics and dissipated power and show they agree with the results given in the main text.
We relate the damping matrix in the two-mode equations of motion to circuit parameters for a dissipator circuit shunted with a resistor.


Consider applying current sources $\vec{I}(t)$ at nodes $a$ and $b$.
Current conservation then leads to the equations of motion 
\ben \hat C \ddot{\vec V}+\hat R^{-1} \dot{\vec V}+\hat L^{-1}\vec V=\dot{\vec{I}}(t);\label{veom0}\een
where  $\vec V=(V_a,V_b)$ and we have defined ($R=R_b$ in this section)
\[\hat R^{-1}=\mqty(\dmat[0]{R_a^{-1},R_b^{-1}})~.\]
where $R_aC_a=\kap^{-1}$ is the bare resonator relaxation time. 
In the absence of damping, these are the Lagrange equations of \eq{lagrange}.
In the frequency domain, \eq{veom0} reads
\ben -i\w\hat Y(\w)\vec V(\w)=-i\w\vec{I}(\w),\een
where we have defined the two port admittance matrix shown in \fig{fig:admit}c
\ben\hat Y(\w)=\hat R^{-1}+i\w\hat C+{\hat L^{-1}}/{ i\w},\label{Yhat}\een
and $\hat C$ and $\hat L$ are given in \eq{CLmatrix}.

\begin{figure}[t]
\includegraphics[width=\linewidth]{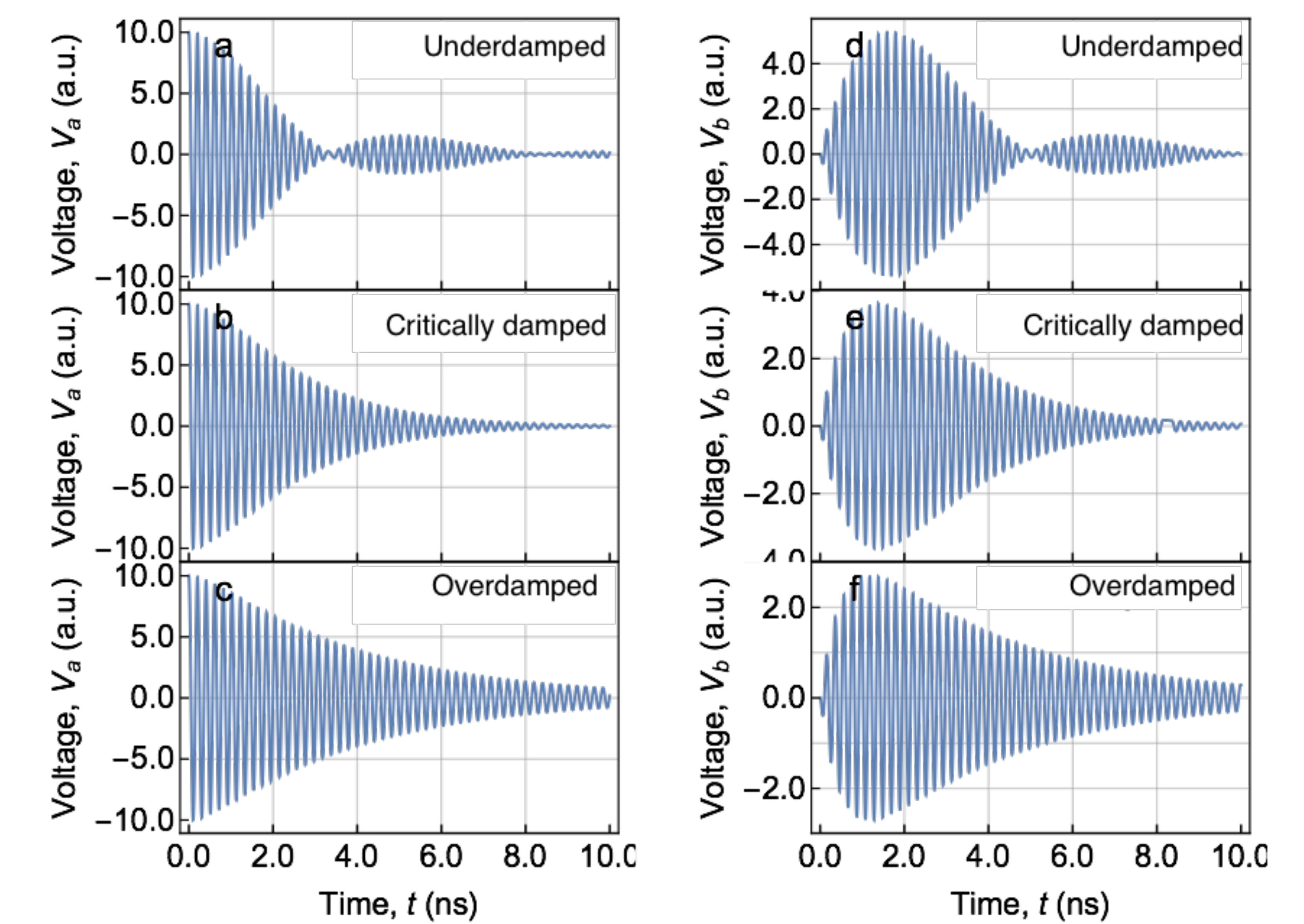}
\caption{Voltages at  nodes $a$ ($V_a$) and $b$ ($V_b$) computed from numerical solution of \eq{veom}, for values of $(\w_0RC_g)^{-1}$ in underdamped ($\w_0RC_g)^{-1}=1$, critically damped ($\w_0RC_g)^{-1}=2$, and overdamped ($\w_0RC_g)^{-1}=3$ regime. }
\label{fig:eom}
\end{figure}

We now consider undriven decay, $\vec I(t)=0$. \eq{veom0} is rewritten as 
\ben \ddot{\vec V}+\hat\g\dot{\vec V}+\hat\W^2 \vec V=0,\label{veom}\een
where 
\begin{subequations} \begin{align} 
\hat\g&\equiv \hat C^{-1}\hat R^{-1}\\
&=\begin{pmatrix}\g_a&\g_{ab}\\\g_{ba}&\g_b\end{pmatrix}
=\begin{pmatrix}
C^{-1}_{aa}C_a\g_a^0&C^{-1}_{ab}C_b\g_b^0\nn
C^{-1}_{ba}C_a\g_a^0&C^{-1}_{bb}C_b\g_b^0\end{pmatrix},\label{gamma}\\
\hat\W^2&=\hat C^{-1}\hat L^{-1}\\
&=\begin{pmatrix}\w^2_a&\w^2_{ab}\\\w^2_{ba}&\w^2_b\end{pmatrix}
=\begin{pmatrix}
C^{-1}_{aa}C_a(\w_a^0)^2&C^{-1}_{ab}C_b(\w_b^0)^2\\
C^{-1}_{ba}C_a(\w_a^0)^2&C^{-1}_{bb}C_b(\w_b^0)^2\end{pmatrix}.\notag
\end{align}\end{subequations}  
where $\g_a^0\equiv{1/R_aC_a}$ and $\g_b^0\equiv{1/R_bC_b}$.
The frequencies $\w_i$ are consistent with \eq{omega}, and the off-diagonal terms in $\hat\W$ are related to the coupling as $\omega _{ab}^2\omega _{ba}^2=(2g)^2(\w_a\w_b)$.
The diagonal terms in the damping tensor are given by
\begin{align}
\g_a&=\frac{1+C_g/C_b}{1+C_g(C_a^{-1}+C_b^{-1})}\frac{1}{R_aC_a}\simeq\frac{1}{R_aC_a}=\kap\nn
\g_b&=\frac{1+C_g/C_a}{1+C_g(C_a^{-1}+C_b^{-1})}\frac{1}{R_bC_b}\simeq\frac{1}{R_bC_b}=\g\label{app:gam}
\end{align}
to $\order{C_g/C_i}$. 
The leading contributions to the off-diagonal terms $\g_{ab}$ are already $\order{C_g/C_i}$.

\begin{figure}[t]
\includegraphics[width=0.7\linewidth]{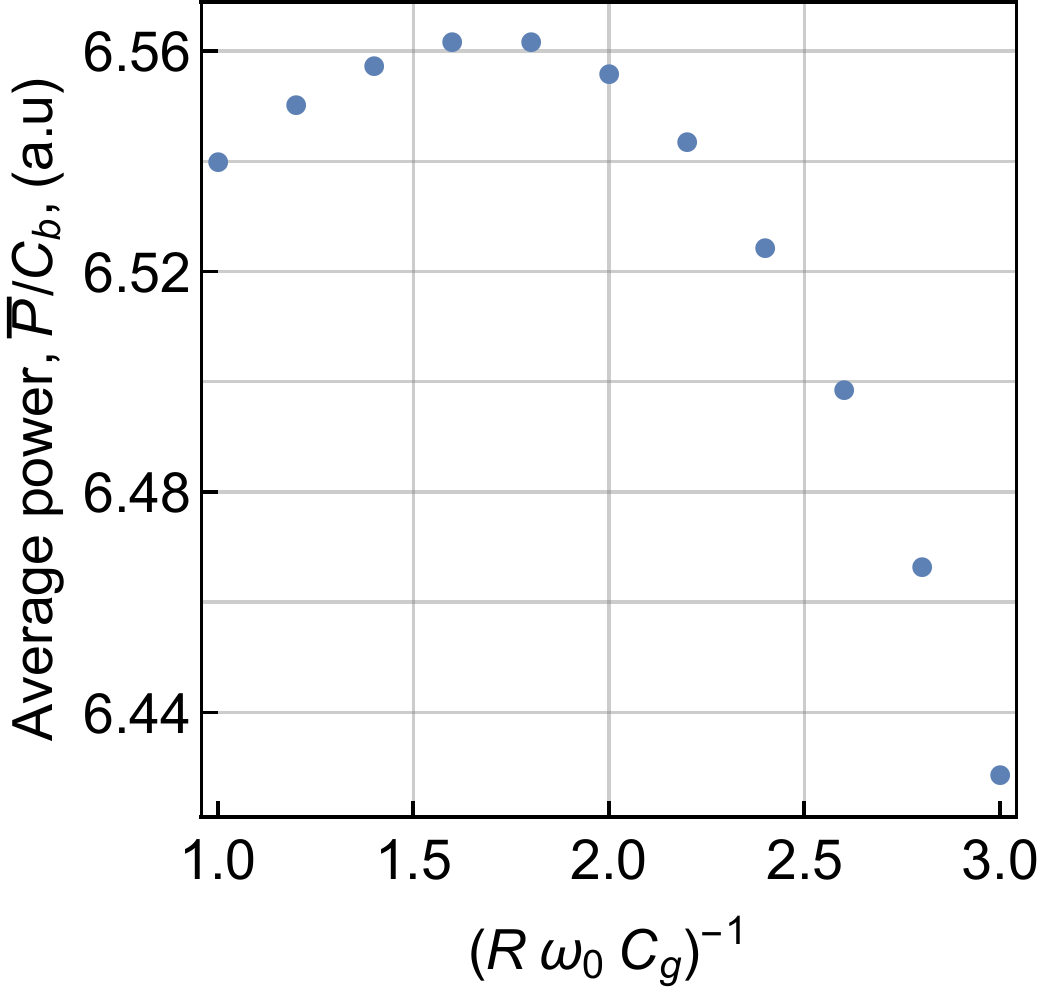}
\caption{Dissipated power  \eq{power} averaged over a period $T=2\pi/g$ as function of $(\w_0RC_g)^{-1}$.}
\label{fig:power}
\end{figure}

\begin{figure}[t]
\includegraphics[width=\linewidth]{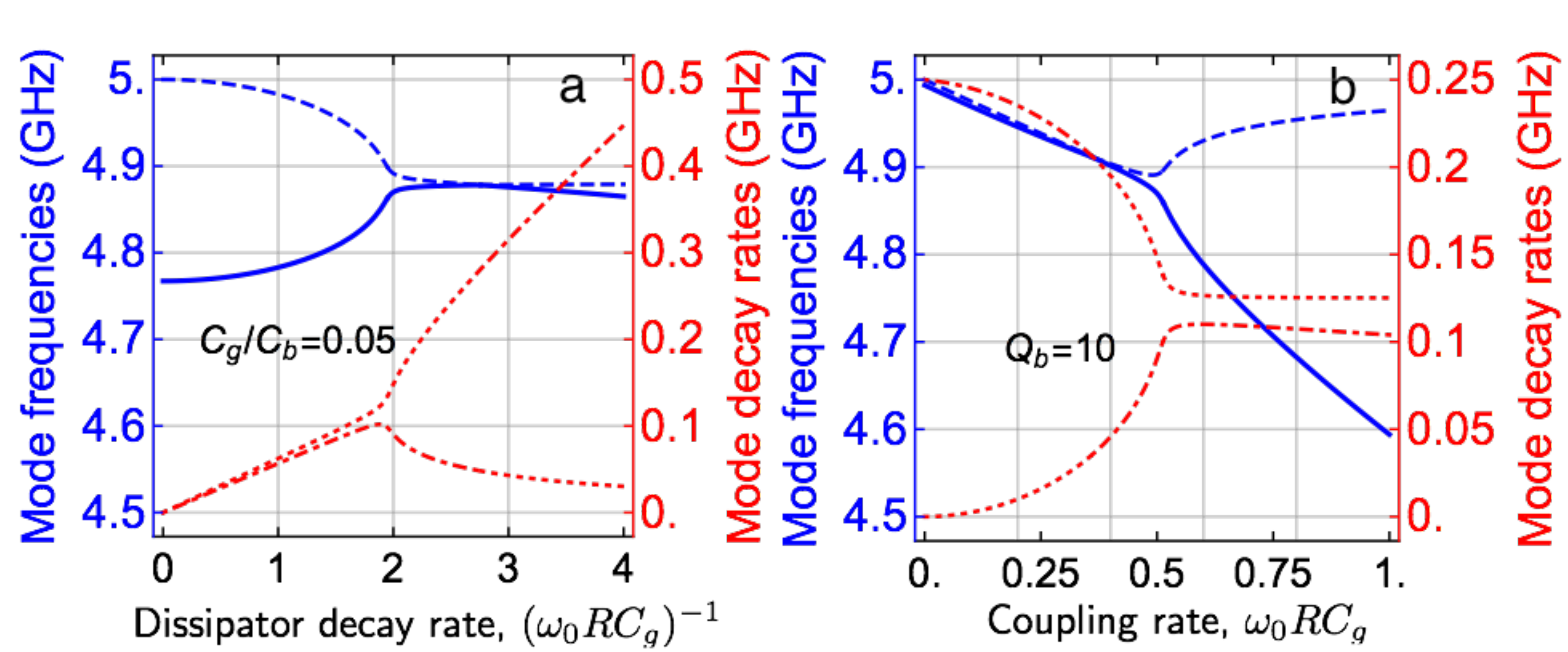}
\caption{Circuit resonant frequencies and damping rates: 
(a) as a function of the dissipator decay rate normalized as $(\w_0RC_g)^{-1}$ at fixed $C_g/C_b=0.05$, and 
(b) as a function of the coupling capacitance normalized as $\w_0RC_g=0.5$ at fixed  dissipator quality factor $Q_b=10$.  
Here, the uncoupled $LC$ frequencies are set on resonance at $\w_0/2\pi=5$ GHz.}
\label{appfig:eig}
\end{figure}

The voltages $V_i(t)$ obtained from the numerical solution of \eq{veom} are plotted in \fig{fig:eom} with initial conditions $V_a(0)=10$, $V_b(0)=0$, $\dot V_a=\dot V_b=0$, for values above, below, and at critical damping  \eq{Rcrit}. 
The plot shows underdamped, critically damped, and overdamped behavior in agreement with the eigenspectrum in \fig{appfig:eig} and \fig{fig:decay} of the main text.  
The dissipated power averaged over a period of inter-mode oscillations $T=2\pi/g$, 
\ben\bar{P}={1\over T}\int_0^Tdt{V_b^2(t)\over R_b},\label{power}\een
is plotted in \fig{fig:power}. 
It shows a maximum as a function of damping slightly below the critical damping point \eq{Rcrit}, in agreement with \fig{fig:T1}  of the main text.

Next, we show that the mode spectrum computed from \eq{veom} agrees with Appendix \ref{app:eom} and \eq{eigval} in the main text. 
The general solution to \eq{veom} reads
\ben\vec V(t)=\Re\sum_n c_n\vec v_n e^{-i\e_n t}~,\quad \Re\e_n>0~,\label{vsol}\een
where the dissipative modes  satisfy
\ben-i\e_n\hat C^{-1}\hat Y(\e_n)\vec v_n=(-\e_n^2-i\e_n\hat\g+\hat\W^2)\vec v_n=0,\een
where $\e_n$ are the complex roots of the characteristic equation
\ben \det(-\w^2-i\w\hat\g+\hat\W^2)=0.\label{det}\een
Since $\e_n\neq0$, the mode vectors $\vec v_n$ are determined by 
\ben\hat Y(\e_n)\vec v_n=0,\een
which implies that 
\ben\det\hat Y(\e_n)=0.\label{detY}\een

Writing $\g_a=\kap$ and taking $\g_{ab}=\g_{ba}=0$, on resonance, the roots of \eq{det} yield
to quadratic order in $\de,\g_b,g$
\ben\w_\pm=\w_0\pm\sqrt{ g^2-\pfrac{\g_-}{4}^2}-i\frac{ \bar\g}{2},\label{delta}\een
consistent to $\order{g/\w_0}$ (see \eq{app:gam}) with \eq{eigval} of the main text.


In the Hamiltonian approach, one works with the coupled equations of motion for charge and flux \eq{Peom} and \eq{Qeom}.
Including the resistor shunt in the dissipator circuit, these equations of motion become
\begin{align}
\vec {\dot{\Phi}}+\hat C^{-1}{\vec Q}=0\label{eq:charge1}\\
\dot{\vec Q}+R^{-1}\dot{\vec{\Phi}}+L^{-1}\vec{\Phi}=0,\notag
\end{align}
which together lead to the EOM for the mode operators
\begin{align}
\begin{pmatrix}\dot a\\\dot b\end{pmatrix}&=\qty[-{i\over2}\til{\g}+\begin{pmatrix}\w_a&-g\\-g&\w_b\end{pmatrix}]\begin{pmatrix}a\\b\end{pmatrix}\nn
&\quad+\qty[{i\over2}\til{\g}+\begin{pmatrix}0&g\\g&0\end{pmatrix}]
\begin{pmatrix}a^\dag\\b^\dag\end{pmatrix}~.\label{app:modeEOM}
\end{align}
Here, we have defined the mode damping matrix
\ben\til{\g}_{ij}=\sqrt{Z_j\over Z_k}\g^T_{jk},\label{gammaH}\een
which differs from $\hat\g$ in \eq{veom} only in the off-diagonal terms of order $\order{g/\w_0}$. 

Neglecting off-diagonal and counter-rotating terms in the damping matrix, the eigenvalues from the secular equation \eq{app:modeEOM} are
\begin{align}
\w_\pm&=\w_0 \sqrt{\qty(1\pm2\sqrt{\frac{g^2-(\g_-/4)^2}{\omega_0^2}})-\qty(\frac{\g_-/4}{\w_0})^2}-i\frac{\bar\gamma}{2}\nn
&\approx\w_0\pm\sqrt{g^2-\pfrac{\g_-}{4}^2}-i\frac{\bar\gamma}{2},
\end{align}
in agreement with \eq{eigval} of the main text.


The resonances defined by \eq{detY} can be related to the zeros of the total admittance $Y_t$ across the resonator by the relation $Y_t=\det\hat Y/Y_b$,  where [see \fig{fig:admit}b]
\ben Y_t(\w )=Y_a(\w )+Y_d(\w ),\label{Ytzeros}\een
and
\ben Y_a=R^{-1}_a+i\w C_a\qty[1-\pfrac{\w_a^0}{\w}^2]\een
is the bare resonator admittance.
In \eq{Ytzeros}, $Y_d$ is admittance of the circuit formed by the coupling capacitor and dissipator, which is given by
\ben Y_d=\frac{Y_gY_b}{Y_b+Y_g},\label{Yd1}\een
where  $Y_g=i\w C_g$ is the coupler admittance, and
\ben Y_b= R^{-1}+i\w C_b\qty[1-\pfrac{\w_b^0}{\w}^2] \een
is the dissipator admittance.  Explicitly,
\begin{subequations}\begin{align}
Y_d
&=\frac{x Q_b C_g (x+i (x^2-1) Q_b)}{R (x^2 Q_b C_g+(x^2-1) C_b Q_b-i x C_b)}\label{admit}
\end{align}\end{subequations}
where we have defined $x=\w/\w_b^0$ and $Q_b=\w_b^0RC_b$ is the dissipator quality factor. 

We denote the complex roots of \eq{Ytzeros} as $\e=\e_r+i\e_i$, $\e_r$ are the resonant mode frequencies and $\e_i$ are mode decay rates.  
The rates $(\e_r,\e_i)/2\pi$ are plotted in \fig{appfig:eig} on resonance $\w_a^0=\w_b^0=\w_0=2\pi\times 5 $ GHz and for $C_a=C_b$.

In  \fig{appfig:eig}a, $\e_r,\e_i$ are plotted as a function of the dissipator decay rate $\g=1/RC_b$ normalized as $(\g/\w_0)(C_b/C_g)=(\w_0RC_g)^{-1}$ for $C_g/C_b=0.05$, which yields $g/2\pi=100$ MHz.  
This plot is qualitatively similar to \fig{appfig:eig}a in the main text. 
The mode decay rate increases as a function of $R^{-1}$ in the underdamped regime up until the  critical damping point at 
\ben R_c^{-1}=2\w_0C_g,\label{Rcrit}\een
which corresponds to $\g_c=4g$.  
At the same point, the two normal mode frequencies come close to merging.

In  \fig{appfig:eig}b, the rates $(\e_r,\e_i)/2\pi$ are plotted as a function of the coupling capacitance normalized as $\w_0RC_g$ at fixed $Q_b=10$ (chosen to cross the curve in  \fig{appfig:eig}a at the critical point where $C_g/C_b=1/2Q_b=0.05$).   
This plot is qualitatively similar to \fig{appfig:eig}b in the main text.
The mode decay rate increases as a function of $C_g$, reaching a maximum at the critical point \eq{Rcrit}, after which it slowly decreases. 

\section{Effective resonator  damping rate including nonlinearity\label{sec:aEOMeff}}
The effective resonator linewidth in the overdamped regime can be controlled by the dissipator frequency.  
To show this, we integrate out $b$ modes along the lines of input-output theory. Using the solution for $b(t)$ in \eq{EOM} in the rotating frame at $\w_a$ and neglecting input fields, we have
\ben b(t)=e^{-z_bt}b(0)+ig\int_0^t\,dt'e^{-z_b(t-t')}a(t')~,\label{eq:b2}\een
where we have defined $z_b=\g/2+i\D_b$. Plugging this into the EOM for $a(t)$ yields
\begin{align} 
\dv{a}{t}&\simeq-\qty[{\kap\over2}-K_a\hat n_a+g^2\int_0^t\,dt'e^{-z_b(t-t')}]a(t)\nn
&-ig e^{-z_bt}b(0)+\order{g^3/\g^2}~,\label{a3}
\end{align}
where in the integral we made the approximation $a(t')\simeq a(t)$ because to $\order{g/\g}$, $a(t)$ decays on the much longer time scale $\kap^{-1}\gg\g^{-1}$.

The EOM for photon number  is
\begin{align}
\dv{\hat{n}_a}{t}&=2{\rm Re}(a^\dag\dot{a})\nn
&\simeq-\kap\hat n_a-2g^2{\rm Re}\int_0^t\,dt'e^{-z_b(t-t')}a^\dag(t)a(t)\nn
&-2g{a^\dag(t)b(0)}e^{-z_bt}\,.
\end{align}
Consider the expectation value in a Fock state $\mel{n_an_b}{\ldots}{n_an_b}$.  Noting that $\ev{a^\dag(t)b(0)}$ vanishes if we assume $b$ is initially in the ground state,
\begin{align}
\ev{\dv{\hat{n}_a}{t}}&\simeq-\qty[\kap+2g^2{\rm Re}\int_0^t\,dt'e^{-{{z_b}}(t-t')}]\ev{n_a(t)}\nn
&=-\qty[\kap+2g^2{\rm Re}\frac{1}{{z_b}}]\ev{n_a(t)}
\end{align}
where we have neglected a transient term that decays with the exponential time scale $\g^{-1}$.  The effective resonator energy decay rate is thus 
\ben\kap'=\kap+\g\frac{g^2}{\g^2/4+\D_b^2}~.\een

\section{Secular approximation in master equation\label{app:master}}

For completeness, we derive here the Lindblad terms used in the density matrix master equation \eq{master}. In particular, we point out that the secular approximation is not valid since we consider strong dissipation, where the dissipator decay rate is of the order of the system energy splittings $g\sim\g$.  In this derivation, we introduce an indexed notation for the system and bath variables $a_1=a$, $a_2=b$, $B_{1\w}=A_\w$, $B_{2\w}=B_\w$.
The total Hamiltonian is $H=H_{\rm sys}+H_{\rm bath}+V$, where
\begin{align}
H_{\rm sys}&=\hbar(\w_1a_1^\dag a_1+\w_2a_2^\dag a_2)+\hbar g(a_1a_2^\dag+a_1^\dag a_2)\nn
H_{\rm bath}&=\hbar \int {d\w\over2\pi} [\w (B^\dag_{1\w} B_{1\w}+B^\dag_{2\w} B_{2\w})\nn
V&=-i\hbar\sum_i\sqrt{\g_i}\int {d\w\over2\pi}(B^\dag_{i\w} a_i-a_i^\dag B_{i\w})~,\label{sysbath}
\end{align}
and where we define $\g_1=\kap$, $\g_2=\g$. Following the standard procedure \cite{gardinerBook04,wallsBook2012}, we go to the interacting picture with respect to $H_{\rm sys}+H_{\rm bath}$: $\rho_I(t)=R^\dag \rho(t)R$, $V_I(t)=R^\dag VR$, where $R=e^{-i(H_{\rm sys}+H_{\rm bath})t}$,
\begin{align}
&V_I=-i\hbar \sum_{m=\pm}[\G_m^\dag(t) c_m-c_m^\dag\G_m(t)]\nn
&\G_m(t)\equiv\int {d\w\over2\pi}C^m_\w e^{-i(\w-\w_m)t}~.
\label{bath}
\end{align} 
Here, we defined the normal mode destruction operators, 
\ben
c_m\equiv U_{mi}a_i,\quad C_{m\w}\equiv\sqrt{\g_i}U_{mi}B_i, \quad U=\begin{pmatrix}\cos\theta&-\sin\theta\\\sin\theta&\cos\theta\end{pmatrix}\label{U},
\een
where $\tan2\theta={2g/\D}$, and $\D=\w_2-\w_1$ is the detuning.  The normal mode frequencies are
\ben\w_\pm=\bar\w\pm\sqrt{g^2+(\D_b/2)^2},\een
where $C_{m\w}$ are the bath modes coupled to $c_m$, and both operators are in the Schrodinger picture,  $R^\dag c_mR=c_me^{-i\w_mt}$.

The density matrix master equation in the interaction picture is 
\begin{align}
\dot\rho_I&=-i[V_I(t),\rho_I(t)]\nn
&\simeq{1\over\hbar^2}\int_0^\infty d\tau [V_I(t),[V_I(t-\tau),\rho_I(t)]],
\label{master1}
\end{align} 
where  $\tau\equiv t-t'$, and we have iterated to second order in $V_I$. In the last line, we took the Markov approximation, $\rho(t-\tau)\to\rho(t)$, and extended the integration limit $t\to\infty$, i.e, $t\gg t_c$, where $t_c$ is the bath correlation time. 
We  assume a factorized density matrix $\rho_I=\rho_{\rm sys}\otimes\rho_{\rm bath}$ and averaged over the bath by tracing over the  thermal bath density matrix $\rho_{\rm bath}=e^{-H_{\rm bath}/T}$; for each  step in the time evolution of $\rho$, the baths are assumed to be at equal temperature $T$. Below we drop the subscript on $\rho_{\rm sys}$, and define the thermal averages
\[\ev{X(t)Y(t')}=\Tr[\rho_{\rm bath}X(t-t')Y(0)];\]
since the bath is stationary the average depends only on $\tau=t-t'$.  For thermal bath, $\ev{\G_m\G_n}=0$ and  $\ev{\G_m^\dag\G_n^\dag}=0$ .
Tracing \eq{master1} over the baths leads to the master equation
\begin{align}
\dot\rho&=\sum_{mn}e^{i\w_{mn}t}{K_{mn}\over2}(2c_m\rho c_n^\dag-c_m^\dag c_n\rho-\rho c_m^\dag c_n)\nn
&+e^{i\w_{mn}t}{G_{mn}\over2}(2c_m^\dag\rho c_n-c_m^\dag c_n\rho-\rho c_m^\dag c_n),
\label{rhodot}
\end{align} 
where $\w_{mn}=\w_m-\w_n$ is  nonzero only for off-diagonal terms $\w_{+-}=\w_+-\w_-=2\sqrt{g^2+(\D_b/2)^2}$. Here, we cannot make the secular approximation, which neglects terms $\sim e^{i\w_{+-}t}K_{+-}$, since we consider strong dissipation so that  the density matrix decay  rate is on the order of the energy splittings $g\simeq \g\sim K_{mn}$. 

The damping rates $K_{mn}$ are set by the frequency space correlators
\begin{align}
&{K_{mn}\over2}\equiv e^{-i\w_{mn}t}\int_0^\infty d\tau\ev{\G_m(t)\G_n^\dag(t-\tau)}\nn
&=\int_0^\infty d\tau e^{-i\w_n\tau}\int{d\w d\w'\over(2\pi)^2}\ev{e^{-i\w\tau}C_{m\w}C^{\dag}_{n\w'}}\nn
&=\sum_i\g_iU_{mi}U_{ni}^*\int_{-\infty}^\infty{d\w\over2\pi}[1+f(\w)]\int_0^\infty d\tau  e^{-i(\w-\w_n)\tau}\nn
&=\sum_i\g_iU_{mi}U_{ni}^*\frac{1+f(\w_m)}{2},
\end{align}
where have we have used the bath correlation functions 
 \begin{align}
 \ev{C_{m\w} C^{\dag}_{n\w'}}&=\sqrt{\g_i\g_j}U_{mi}U_{nj}^*\ev{B_{i\w} B^\dag_{j\w'}}\nn
 \ev{B_{i\w} B^\dag_{j\w'}}&=2\pi\de_{ij}\de(\w-\w')(1+f(\w))
 \end{align}
and where  $f(\w)$ is the Bose function. Here, we have used the identity
\[\int_0^\infty d\tau  e^{-i(\w-\w_n)\tau}=\pi\de(\w-\w_n)+i\mcal{P}{1\over\w-\w_n}\]
and neglected the imaginary parts which cause  Stark shifts. 
Explicitly, 
\begin{align}
K_{++}&=(\kap\cos^2\theta+\g\sin^2\theta)(1+f(\w_+))\nn
K_{--}&=(\kap\sin^2\theta+\g\cos^2\theta)(1+f(\w_-))\nn
K_{\pm\mp}&=(\kap-\g)\cos\theta\sin\theta(1+f(\w_\mp))~.
\end{align}
Similarly
\begin{align}
{G_{mn}\over2}&\equiv e^{-i\w_{mn}t}\int_0^\infty d\tau\ev{\G^\dag_m(t)\G_n(t-\tau)}\nn
&=\sum_i\g_iU_{mi}U_{ni}^*\frac{f(\w_n)}{2}~.
\end{align}
At low temperatures $f(\w_n)$ is negligible, so are the absorption terms which go as $G_{mn}\sim f(\w_n)$.  
Transforming \eq{rhodot} back to the Schrodinger picture by $\dot\rho\to U\dot\rho U^\dag+i[H_{\rm sys},\rho]$, we find
\begin{align}
\dot\rho&=-i[H_{\rm sys},\rho]+\sum_{mn}{K_{mn}\over2}(2c_m\rho c_n^\dag-c_m^\dag c_n\rho-\rho c_m^\dag c_n)\nn
&=-i[H_{\rm sys},\rho]+\sum_i\g_i\mcal D(a_i)\label{master3}
\end{align} 
where  $\mcal{D}(x)\rho=(2x\rho x^\dag-x^\dag x\rho-\rho x^\dag x)/2$ is the Linblad superoperator, and we have used the identity
\begin{align}
K_{mn}c_m^\dag c_n&=\sum_i\g_i(c_m^\dag U_{mi})(U_{ni}^* c_n)\nn
&=\sum_i\g_i(U^\dag_{im}c_m)^\dag(U^\dag_{in} c_n)=\g_ia_i^\dag a_i~.
\end{align}
It is thus appropriate to use the decoupled Lindbladians in the master equation 
\eq{master}.

\bibliography{physics}

\begin{thebibliography}{50}%
\makeatletter
\providecommand \@ifxundefined [1]{%
 \@ifx{#1\undefined}
}%
\providecommand \@ifnum [1]{%
 \ifnum #1\expandafter \@firstoftwo
 \else \expandafter \@secondoftwo
 \fi
}%
\providecommand \@ifx [1]{%
 \ifx #1\expandafter \@firstoftwo
 \else \expandafter \@secondoftwo
 \fi
}%
\providecommand \natexlab [1]{#1}%
\providecommand \enquote  [1]{``#1''}%
\providecommand \bibnamefont  [1]{#1}%
\providecommand \bibfnamefont [1]{#1}%
\providecommand \citenamefont [1]{#1}%
\providecommand \href@noop [0]{\@secondoftwo}%
\providecommand \href [0]{\begingroup \@sanitize@url \@href}%
\providecommand \@href[1]{\@@startlink{#1}\@@href}%
\providecommand \@@href[1]{\endgroup#1\@@endlink}%
\providecommand \@sanitize@url [0]{\catcode `\\12\catcode `\$12\catcode
  `\&12\catcode `\#12\catcode `\^12\catcode `\_12\catcode `\%12\relax}%
\providecommand \@@startlink[1]{}%
\providecommand \@@endlink[0]{}%
\providecommand \url  [0]{\begingroup\@sanitize@url \@url }%
\providecommand \@url [1]{\endgroup\@href {#1}{\urlprefix }}%
\providecommand \urlprefix  [0]{URL }%
\providecommand \Eprint [0]{\href }%
\providecommand \doibase [0]{http://dx.doi.org/}%
\providecommand \selectlanguage [0]{\@gobble}%
\providecommand \bibinfo  [0]{\@secondoftwo}%
\providecommand \bibfield  [0]{\@secondoftwo}%
\providecommand \translation [1]{[#1]}%
\providecommand \BibitemOpen [0]{}%
\providecommand \bibitemStop [0]{}%
\providecommand \bibitemNoStop [0]{.\EOS\space}%
\providecommand \EOS [0]{\spacefactor3000\relax}%
\providecommand \BibitemShut  [1]{\csname bibitem#1\endcsname}%
\let\auto@bib@innerbib\@empty
\bibitem [{\citenamefont {Fowler}\ \emph {et~al.}(2012)\citenamefont {Fowler},
  \citenamefont {Mariantoni}, \citenamefont {Martinis},\ and\ \citenamefont
  {Cleland}}]{fowlerPRA12}%
  \BibitemOpen
  \bibfield  {author} {\bibinfo {author} {\bibfnamefont {A.~G.}\ \bibnamefont
  {Fowler}}, \bibinfo {author} {\bibfnamefont {M.}~\bibnamefont {Mariantoni}},
  \bibinfo {author} {\bibfnamefont {J.~M.}\ \bibnamefont {Martinis}}, \ and\
  \bibinfo {author} {\bibfnamefont {A.~N.}\ \bibnamefont {Cleland}},\ }\href
  {\doibase 10.1103/PhysRevA.86.032324} {\bibfield  {journal} {\bibinfo
  {journal} {Phys. Rev. A}\ }\textbf {\bibinfo {volume} {86}},\ \bibinfo
  {pages} {032324} (\bibinfo {year} {2012})}\BibitemShut {NoStop}%
\bibitem [{\citenamefont {Kelly}\ \emph {et~al.}(2015)\citenamefont {Kelly},
  \citenamefont {Barends}, \citenamefont {Fowler}, \citenamefont {Megrant},
  \citenamefont {Jeffrey}, \citenamefont {White}, \citenamefont {Sank},
  \citenamefont {Mutus}, \citenamefont {Campbell}, \citenamefont {Chen},
  \citenamefont {Chen}, \citenamefont {Chiaro}, \citenamefont {Dunsworth},
  \citenamefont {Hoi}, \citenamefont {Neill}, \citenamefont {O'Malley},
  \citenamefont {Quintana}, \citenamefont {Roushan}, \citenamefont
  {Vainsencher}, \citenamefont {Wenner}, \citenamefont {Cleland},\ and\
  \citenamefont {Martinis}}]{kellyNAT15}%
  \BibitemOpen
  \bibfield  {author} {\bibinfo {author} {\bibfnamefont {J.}~\bibnamefont
  {Kelly}}, \bibinfo {author} {\bibfnamefont {R.}~\bibnamefont {Barends}},
  \bibinfo {author} {\bibfnamefont {A.~G.}\ \bibnamefont {Fowler}}, \bibinfo
  {author} {\bibfnamefont {A.}~\bibnamefont {Megrant}}, \bibinfo {author}
  {\bibfnamefont {E.}~\bibnamefont {Jeffrey}}, \bibinfo {author} {\bibfnamefont
  {T.~C.}\ \bibnamefont {White}}, \bibinfo {author} {\bibfnamefont
  {D.}~\bibnamefont {Sank}}, \bibinfo {author} {\bibfnamefont {J.~Y.}\
  \bibnamefont {Mutus}}, \bibinfo {author} {\bibfnamefont {B.}~\bibnamefont
  {Campbell}}, \bibinfo {author} {\bibfnamefont {Y.}~\bibnamefont {Chen}},
  \bibinfo {author} {\bibfnamefont {Z.}~\bibnamefont {Chen}}, \bibinfo {author}
  {\bibfnamefont {B.}~\bibnamefont {Chiaro}}, \bibinfo {author} {\bibfnamefont
  {A.}~\bibnamefont {Dunsworth}}, \bibinfo {author} {\bibfnamefont {I.~C.}\
  \bibnamefont {Hoi}}, \bibinfo {author} {\bibfnamefont {C.}~\bibnamefont
  {Neill}}, \bibinfo {author} {\bibfnamefont {P.~J.~J.}\ \bibnamefont
  {O'Malley}}, \bibinfo {author} {\bibfnamefont {C.}~\bibnamefont {Quintana}},
  \bibinfo {author} {\bibfnamefont {P.}~\bibnamefont {Roushan}}, \bibinfo
  {author} {\bibfnamefont {A.}~\bibnamefont {Vainsencher}}, \bibinfo {author}
  {\bibfnamefont {J.}~\bibnamefont {Wenner}}, \bibinfo {author} {\bibfnamefont
  {A.~N.}\ \bibnamefont {Cleland}}, \ and\ \bibinfo {author} {\bibfnamefont
  {J.~M.}\ \bibnamefont {Martinis}},\ }\href
  {http://dx.doi.org/10.1038/nature14270} {\bibfield  {journal} {\bibinfo
  {journal} {Nature}\ }\textbf {\bibinfo {volume} {519}},\ \bibinfo {pages}
  {66} (\bibinfo {year} {2015})}\BibitemShut {NoStop}%
\bibitem [{\citenamefont {Ofek}\ \emph {et~al.}(2016)\citenamefont {Ofek},
  \citenamefont {Petrenko}, \citenamefont {Heeres}, \citenamefont {Reinhold},
  \citenamefont {Leghtas}, \citenamefont {Vlastakis}, \citenamefont {Liu},
  \citenamefont {Frunzio}, \citenamefont {Girvin}, \citenamefont {Jiang},
  \citenamefont {Mirrahimi}, \citenamefont {Devoret},\ and\ \citenamefont
  {Schoelkopf}}]{ofekNAT16}%
  \BibitemOpen
  \bibfield  {author} {\bibinfo {author} {\bibfnamefont {N.}~\bibnamefont
  {Ofek}}, \bibinfo {author} {\bibfnamefont {A.}~\bibnamefont {Petrenko}},
  \bibinfo {author} {\bibfnamefont {R.}~\bibnamefont {Heeres}}, \bibinfo
  {author} {\bibfnamefont {P.}~\bibnamefont {Reinhold}}, \bibinfo {author}
  {\bibfnamefont {Z.}~\bibnamefont {Leghtas}}, \bibinfo {author} {\bibfnamefont
  {B.}~\bibnamefont {Vlastakis}}, \bibinfo {author} {\bibfnamefont
  {Y.}~\bibnamefont {Liu}}, \bibinfo {author} {\bibfnamefont {L.}~\bibnamefont
  {Frunzio}}, \bibinfo {author} {\bibfnamefont {S.~M.}\ \bibnamefont {Girvin}},
  \bibinfo {author} {\bibfnamefont {L.}~\bibnamefont {Jiang}}, \bibinfo
  {author} {\bibfnamefont {M.}~\bibnamefont {Mirrahimi}}, \bibinfo {author}
  {\bibfnamefont {M.~H.}\ \bibnamefont {Devoret}}, \ and\ \bibinfo {author}
  {\bibfnamefont {R.~J.}\ \bibnamefont {Schoelkopf}},\ }\href
  {http://dx.doi.org/10.1038/nature18949} {\bibfield  {journal} {\bibinfo
  {journal} {Nature}\ }\textbf {\bibinfo {volume} {536}},\ \bibinfo {pages}
  {441 EP } (\bibinfo {year} {2016})}\BibitemShut {NoStop}%
\bibitem [{\citenamefont {Gu}\ \emph {et~al.}(2017)\citenamefont {Gu},
  \citenamefont {Kockum}, \citenamefont {Miranowicz}, \citenamefont {xi~Liu},\
  and\ \citenamefont {Nori}}]{guPR17}%
  \BibitemOpen
  \bibfield  {author} {\bibinfo {author} {\bibfnamefont {X.}~\bibnamefont
  {Gu}}, \bibinfo {author} {\bibfnamefont {A.~F.}\ \bibnamefont {Kockum}},
  \bibinfo {author} {\bibfnamefont {A.}~\bibnamefont {Miranowicz}}, \bibinfo
  {author} {\bibfnamefont {Y.}~\bibnamefont {xi~Liu}}, \ and\ \bibinfo {author}
  {\bibfnamefont {F.}~\bibnamefont {Nori}},\ }\href {\doibase
  https://doi.org/10.1016/j.physrep.2017.10.002} {\bibfield  {journal}
  {\bibinfo  {journal} {Physics Reports}\ }\textbf {\bibinfo {volume}
  {718-719}},\ \bibinfo {pages} {1 } (\bibinfo {year} {2017})}\BibitemShut
  {NoStop}%
\bibitem [{\citenamefont {Gambetta}\ \emph {et~al.}(2006)\citenamefont
  {Gambetta}, \citenamefont {Blais}, \citenamefont {Schuster}, \citenamefont
  {Wallraff}, \citenamefont {Frunzio}, \citenamefont {Majer}, \citenamefont
  {Devoret}, \citenamefont {Girvin},\ and\ \citenamefont
  {Schoelkopf}}]{gambettaPRA06}%
  \BibitemOpen
  \bibfield  {author} {\bibinfo {author} {\bibfnamefont {J.}~\bibnamefont
  {Gambetta}}, \bibinfo {author} {\bibfnamefont {A.}~\bibnamefont {Blais}},
  \bibinfo {author} {\bibfnamefont {D.~I.}\ \bibnamefont {Schuster}}, \bibinfo
  {author} {\bibfnamefont {A.}~\bibnamefont {Wallraff}}, \bibinfo {author}
  {\bibfnamefont {L.}~\bibnamefont {Frunzio}}, \bibinfo {author} {\bibfnamefont
  {J.}~\bibnamefont {Majer}}, \bibinfo {author} {\bibfnamefont {M.~H.}\
  \bibnamefont {Devoret}}, \bibinfo {author} {\bibfnamefont {S.~M.}\
  \bibnamefont {Girvin}}, \ and\ \bibinfo {author} {\bibfnamefont {R.~J.}\
  \bibnamefont {Schoelkopf}},\ }\href {\doibase 10.1103/PhysRevA.74.042318}
  {\bibfield  {journal} {\bibinfo  {journal} {Phys. Rev. A}\ }\textbf {\bibinfo
  {volume} {74}},\ \bibinfo {pages} {042318} (\bibinfo {year}
  {2006})}\BibitemShut {NoStop}%
\bibitem [{\citenamefont {Kandala}\ \emph {et~al.}(2017)\citenamefont
  {Kandala}, \citenamefont {Mezzacapo}, \citenamefont {Temme}, \citenamefont
  {Takita}, \citenamefont {Brink}, \citenamefont {Chow},\ and\ \citenamefont
  {Gambetta}}]{kandalaNAT17}%
  \BibitemOpen
  \bibfield  {author} {\bibinfo {author} {\bibfnamefont {A.}~\bibnamefont
  {Kandala}}, \bibinfo {author} {\bibfnamefont {A.}~\bibnamefont {Mezzacapo}},
  \bibinfo {author} {\bibfnamefont {K.}~\bibnamefont {Temme}}, \bibinfo
  {author} {\bibfnamefont {M.}~\bibnamefont {Takita}}, \bibinfo {author}
  {\bibfnamefont {M.}~\bibnamefont {Brink}}, \bibinfo {author} {\bibfnamefont
  {J.~M.}\ \bibnamefont {Chow}}, \ and\ \bibinfo {author} {\bibfnamefont
  {J.~M.}\ \bibnamefont {Gambetta}},\ }\href
  {http://dx.doi.org/10.1038/nature23879} {\bibfield  {journal} {\bibinfo
  {journal} {Nature}\ }\textbf {\bibinfo {volume} {549}},\ \bibinfo {pages}
  {242 EP } (\bibinfo {year} {2017})}\BibitemShut {NoStop}%
\bibitem [{\citenamefont {Georgescu}\ \emph {et~al.}(2014)\citenamefont
  {Georgescu}, \citenamefont {Ashhab},\ and\ \citenamefont
  {Nori}}]{georgescuRMP14}%
  \BibitemOpen
  \bibfield  {author} {\bibinfo {author} {\bibfnamefont {I.~M.}\ \bibnamefont
  {Georgescu}}, \bibinfo {author} {\bibfnamefont {S.}~\bibnamefont {Ashhab}}, \
  and\ \bibinfo {author} {\bibfnamefont {F.}~\bibnamefont {Nori}},\ }\href
  {\doibase 10.1103/RevModPhys.86.153} {\bibfield  {journal} {\bibinfo
  {journal} {Rev. Mod. Phys.}\ }\textbf {\bibinfo {volume} {86}},\ \bibinfo
  {pages} {153} (\bibinfo {year} {2014})}\BibitemShut {NoStop}%
\bibitem [{\citenamefont {Reed}\ \emph {et~al.}(2010)\citenamefont {Reed},
  \citenamefont {Johnson}, \citenamefont {Houck}, \citenamefont {DiCarlo},
  \citenamefont {Chow}, \citenamefont {Schuster}, \citenamefont {Frunzio},\
  and\ \citenamefont {Schoelkopf}}]{reedAPL10}%
  \BibitemOpen
  \bibfield  {author} {\bibinfo {author} {\bibfnamefont {M.~D.}\ \bibnamefont
  {Reed}}, \bibinfo {author} {\bibfnamefont {B.~R.}\ \bibnamefont {Johnson}},
  \bibinfo {author} {\bibfnamefont {A.~A.}\ \bibnamefont {Houck}}, \bibinfo
  {author} {\bibfnamefont {L.}~\bibnamefont {DiCarlo}}, \bibinfo {author}
  {\bibfnamefont {J.~M.}\ \bibnamefont {Chow}}, \bibinfo {author}
  {\bibfnamefont {D.~I.}\ \bibnamefont {Schuster}}, \bibinfo {author}
  {\bibfnamefont {L.}~\bibnamefont {Frunzio}}, \ and\ \bibinfo {author}
  {\bibfnamefont {R.~J.}\ \bibnamefont {Schoelkopf}},\ }\href {\doibase
  10.1063/1.3435463} {\bibfield  {journal} {\bibinfo  {journal} {Applied
  Physics Letters}\ }\textbf {\bibinfo {volume} {96}},\ \bibinfo {pages}
  {203110} (\bibinfo {year} {2010})}\BibitemShut {NoStop}%
\bibitem [{\citenamefont {Jeffrey}\ \emph {et~al.}(2014)\citenamefont
  {Jeffrey}, \citenamefont {Sank}, \citenamefont {Mutus}, \citenamefont
  {White}, \citenamefont {Kelly}, \citenamefont {Barends}, \citenamefont
  {Chen}, \citenamefont {Chen}, \citenamefont {Chiaro}, \citenamefont
  {Dunsworth}, \citenamefont {Megrant}, \citenamefont {O'Malley}, \citenamefont
  {Neill}, \citenamefont {Roushan}, \citenamefont {Vainsencher}, \citenamefont
  {Wenner}, \citenamefont {Cleland},\ and\ \citenamefont
  {Martinis}}]{jeffreyPRL14}%
  \BibitemOpen
  \bibfield  {author} {\bibinfo {author} {\bibfnamefont {E.}~\bibnamefont
  {Jeffrey}}, \bibinfo {author} {\bibfnamefont {D.}~\bibnamefont {Sank}},
  \bibinfo {author} {\bibfnamefont {J.~Y.}\ \bibnamefont {Mutus}}, \bibinfo
  {author} {\bibfnamefont {T.~C.}\ \bibnamefont {White}}, \bibinfo {author}
  {\bibfnamefont {J.}~\bibnamefont {Kelly}}, \bibinfo {author} {\bibfnamefont
  {R.}~\bibnamefont {Barends}}, \bibinfo {author} {\bibfnamefont
  {Y.}~\bibnamefont {Chen}}, \bibinfo {author} {\bibfnamefont {Z.}~\bibnamefont
  {Chen}}, \bibinfo {author} {\bibfnamefont {B.}~\bibnamefont {Chiaro}},
  \bibinfo {author} {\bibfnamefont {A.}~\bibnamefont {Dunsworth}}, \bibinfo
  {author} {\bibfnamefont {A.}~\bibnamefont {Megrant}}, \bibinfo {author}
  {\bibfnamefont {P.~J.~J.}\ \bibnamefont {O'Malley}}, \bibinfo {author}
  {\bibfnamefont {C.}~\bibnamefont {Neill}}, \bibinfo {author} {\bibfnamefont
  {P.}~\bibnamefont {Roushan}}, \bibinfo {author} {\bibfnamefont
  {A.}~\bibnamefont {Vainsencher}}, \bibinfo {author} {\bibfnamefont
  {J.}~\bibnamefont {Wenner}}, \bibinfo {author} {\bibfnamefont {A.~N.}\
  \bibnamefont {Cleland}}, \ and\ \bibinfo {author} {\bibfnamefont {J.~M.}\
  \bibnamefont {Martinis}},\ }\href {\doibase 10.1103/PhysRevLett.112.190504}
  {\bibfield  {journal} {\bibinfo  {journal} {Phys. Rev. Lett.}\ }\textbf
  {\bibinfo {volume} {112}},\ \bibinfo {pages} {190504} (\bibinfo {year}
  {2014})}\BibitemShut {NoStop}%
\bibitem [{\citenamefont {Yan}\ \emph {et~al.}(2016)\citenamefont {Yan},
  \citenamefont {Gustavsson}, \citenamefont {Kamal}, \citenamefont {Birenbaum},
  \citenamefont {Sears}, \citenamefont {Hover}, \citenamefont {Gudmundsen},
  \citenamefont {Rosenberg}, \citenamefont {Samach}, \citenamefont {Weber},
  \citenamefont {Yoder}, \citenamefont {Orlando}, \citenamefont {Clarke},
  \citenamefont {Kerman},\ and\ \citenamefont {Oliver}}]{yanNATC16}%
  \BibitemOpen
  \bibfield  {author} {\bibinfo {author} {\bibfnamefont {F.}~\bibnamefont
  {Yan}}, \bibinfo {author} {\bibfnamefont {S.}~\bibnamefont {Gustavsson}},
  \bibinfo {author} {\bibfnamefont {A.}~\bibnamefont {Kamal}}, \bibinfo
  {author} {\bibfnamefont {J.}~\bibnamefont {Birenbaum}}, \bibinfo {author}
  {\bibfnamefont {A.~P.}\ \bibnamefont {Sears}}, \bibinfo {author}
  {\bibfnamefont {D.}~\bibnamefont {Hover}}, \bibinfo {author} {\bibfnamefont
  {T.~J.}\ \bibnamefont {Gudmundsen}}, \bibinfo {author} {\bibfnamefont
  {D.}~\bibnamefont {Rosenberg}}, \bibinfo {author} {\bibfnamefont
  {G.}~\bibnamefont {Samach}}, \bibinfo {author} {\bibfnamefont
  {S.}~\bibnamefont {Weber}}, \bibinfo {author} {\bibfnamefont {J.~L.}\
  \bibnamefont {Yoder}}, \bibinfo {author} {\bibfnamefont {T.~P.}\ \bibnamefont
  {Orlando}}, \bibinfo {author} {\bibfnamefont {J.}~\bibnamefont {Clarke}},
  \bibinfo {author} {\bibfnamefont {A.~J.}\ \bibnamefont {Kerman}}, \ and\
  \bibinfo {author} {\bibfnamefont {W.~D.}\ \bibnamefont {Oliver}},\ }\href
  {http://dx.doi.org/10.1038/ncomms12964} {\bibfield  {journal} {\bibinfo
  {journal} {Nature Communications}\ }\textbf {\bibinfo {volume} {7}},\
  \bibinfo {pages} {12964 EP } (\bibinfo {year} {2016})}\BibitemShut {NoStop}%
\bibitem [{\citenamefont {McClure}\ \emph {et~al.}(2016)\citenamefont
  {McClure}, \citenamefont {Paik}, \citenamefont {Bishop}, \citenamefont
  {Steffen}, \citenamefont {Chow},\ and\ \citenamefont
  {Gambetta}}]{mcclurePRA16}%
  \BibitemOpen
  \bibfield  {author} {\bibinfo {author} {\bibfnamefont {D.~T.}\ \bibnamefont
  {McClure}}, \bibinfo {author} {\bibfnamefont {H.}~\bibnamefont {Paik}},
  \bibinfo {author} {\bibfnamefont {L.~S.}\ \bibnamefont {Bishop}}, \bibinfo
  {author} {\bibfnamefont {M.}~\bibnamefont {Steffen}}, \bibinfo {author}
  {\bibfnamefont {J.~M.}\ \bibnamefont {Chow}}, \ and\ \bibinfo {author}
  {\bibfnamefont {J.~M.}\ \bibnamefont {Gambetta}},\ }\href {\doibase
  10.1103/PhysRevApplied.5.011001} {\bibfield  {journal} {\bibinfo  {journal}
  {Phys. Rev. Applied}\ }\textbf {\bibinfo {volume} {5}},\ \bibinfo {pages}
  {011001} (\bibinfo {year} {2016})}\BibitemShut {NoStop}%
\bibitem [{\citenamefont {Bultink}\ \emph {et~al.}(2016)\citenamefont
  {Bultink}, \citenamefont {Rol}, \citenamefont {O'Brien}, \citenamefont {Fu},
  \citenamefont {Dikken}, \citenamefont {Dickel}, \citenamefont {Vermeulen},
  \citenamefont {de~Sterke}, \citenamefont {Bruno}, \citenamefont {Schouten},\
  and\ \citenamefont {DiCarlo}}]{bultinkPRA16}%
  \BibitemOpen
  \bibfield  {author} {\bibinfo {author} {\bibfnamefont {C.~C.}\ \bibnamefont
  {Bultink}}, \bibinfo {author} {\bibfnamefont {M.~A.}\ \bibnamefont {Rol}},
  \bibinfo {author} {\bibfnamefont {T.~E.}\ \bibnamefont {O'Brien}}, \bibinfo
  {author} {\bibfnamefont {X.}~\bibnamefont {Fu}}, \bibinfo {author}
  {\bibfnamefont {B.~C.~S.}\ \bibnamefont {Dikken}}, \bibinfo {author}
  {\bibfnamefont {C.}~\bibnamefont {Dickel}}, \bibinfo {author} {\bibfnamefont
  {R.~F.~L.}\ \bibnamefont {Vermeulen}}, \bibinfo {author} {\bibfnamefont
  {J.~C.}\ \bibnamefont {de~Sterke}}, \bibinfo {author} {\bibfnamefont
  {A.}~\bibnamefont {Bruno}}, \bibinfo {author} {\bibfnamefont {R.~N.}\
  \bibnamefont {Schouten}}, \ and\ \bibinfo {author} {\bibfnamefont
  {L.}~\bibnamefont {DiCarlo}},\ }\href {\doibase
  10.1103/PhysRevApplied.6.034008} {\bibfield  {journal} {\bibinfo  {journal}
  {Phys. Rev. Applied}\ }\textbf {\bibinfo {volume} {6}},\ \bibinfo {pages}
  {034008} (\bibinfo {year} {2016})}\BibitemShut {NoStop}%
\bibitem [{\citenamefont {Boutin}\ \emph {et~al.}(2017)\citenamefont {Boutin},
  \citenamefont {Andersen}, \citenamefont {Venkatraman}, \citenamefont
  {Ferris},\ and\ \citenamefont {Blais}}]{boutinPRA17}%
  \BibitemOpen
  \bibfield  {author} {\bibinfo {author} {\bibfnamefont {S.}~\bibnamefont
  {Boutin}}, \bibinfo {author} {\bibfnamefont {C.~K.}\ \bibnamefont
  {Andersen}}, \bibinfo {author} {\bibfnamefont {J.}~\bibnamefont
  {Venkatraman}}, \bibinfo {author} {\bibfnamefont {A.~J.}\ \bibnamefont
  {Ferris}}, \ and\ \bibinfo {author} {\bibfnamefont {A.}~\bibnamefont
  {Blais}},\ }\href {\doibase 10.1103/PhysRevA.96.042315} {\bibfield  {journal}
  {\bibinfo  {journal} {Phys. Rev. A}\ }\textbf {\bibinfo {volume} {96}},\
  \bibinfo {pages} {042315} (\bibinfo {year} {2017})}\BibitemShut {NoStop}%
\bibitem [{\citenamefont {Opremcak}\ \emph {et~al.}(2018)\citenamefont
  {Opremcak}, \citenamefont {Pechenezhskiy}, \citenamefont {Howington},
  \citenamefont {Christensen}, \citenamefont {Beck}, \citenamefont {Leonard},
  \citenamefont {Suttle}, \citenamefont {Wilen}, \citenamefont {Nesterov},
  \citenamefont {Ribeill}, \citenamefont {Thorbeck}, \citenamefont {Schlenker},
  \citenamefont {Vavilov}, \citenamefont {Plourde},\ and\ \citenamefont
  {McDermott}}]{opremcakSCI18}%
  \BibitemOpen
  \bibfield  {author} {\bibinfo {author} {\bibfnamefont {A.}~\bibnamefont
  {Opremcak}}, \bibinfo {author} {\bibfnamefont {I.~V.}\ \bibnamefont
  {Pechenezhskiy}}, \bibinfo {author} {\bibfnamefont {C.}~\bibnamefont
  {Howington}}, \bibinfo {author} {\bibfnamefont {B.~G.}\ \bibnamefont
  {Christensen}}, \bibinfo {author} {\bibfnamefont {M.~A.}\ \bibnamefont
  {Beck}}, \bibinfo {author} {\bibfnamefont {E.}~\bibnamefont {Leonard}},
  \bibinfo {author} {\bibfnamefont {J.}~\bibnamefont {Suttle}}, \bibinfo
  {author} {\bibfnamefont {C.}~\bibnamefont {Wilen}}, \bibinfo {author}
  {\bibfnamefont {K.~N.}\ \bibnamefont {Nesterov}}, \bibinfo {author}
  {\bibfnamefont {G.~J.}\ \bibnamefont {Ribeill}}, \bibinfo {author}
  {\bibfnamefont {T.}~\bibnamefont {Thorbeck}}, \bibinfo {author}
  {\bibfnamefont {F.}~\bibnamefont {Schlenker}}, \bibinfo {author}
  {\bibfnamefont {M.~G.}\ \bibnamefont {Vavilov}}, \bibinfo {author}
  {\bibfnamefont {B.~L.~T.}\ \bibnamefont {Plourde}}, \ and\ \bibinfo {author}
  {\bibfnamefont {R.}~\bibnamefont {McDermott}},\ }\href {\doibase
  10.1126/science.aat4625} {\bibfield  {journal} {\bibinfo  {journal}
  {Science}\ }\textbf {\bibinfo {volume} {361}},\ \bibinfo {pages} {1239}
  (\bibinfo {year} {2018})}\BibitemShut {NoStop}%
\bibitem [{\citenamefont {Partanen}\ \emph {et~al.}(2018)\citenamefont
  {Partanen}, \citenamefont {Tan}, \citenamefont {Masuda}, \citenamefont
  {Govenius}, \citenamefont {Lake}, \citenamefont {Jenei}, \citenamefont
  {Gr{\"o}nberg}, \citenamefont {Hassel}, \citenamefont {Simbierowicz},
  \citenamefont {Vesterinen}, \citenamefont {Tuorila}, \citenamefont
  {Ala-Nissila},\ and\ \citenamefont {M{\"o}tt{\"o}nen}}]{partanenSR18}%
  \BibitemOpen
  \bibfield  {author} {\bibinfo {author} {\bibfnamefont {M.}~\bibnamefont
  {Partanen}}, \bibinfo {author} {\bibfnamefont {K.~Y.}\ \bibnamefont {Tan}},
  \bibinfo {author} {\bibfnamefont {S.}~\bibnamefont {Masuda}}, \bibinfo
  {author} {\bibfnamefont {J.}~\bibnamefont {Govenius}}, \bibinfo {author}
  {\bibfnamefont {R.~E.}\ \bibnamefont {Lake}}, \bibinfo {author}
  {\bibfnamefont {M.}~\bibnamefont {Jenei}}, \bibinfo {author} {\bibfnamefont
  {L.}~\bibnamefont {Gr{\"o}nberg}}, \bibinfo {author} {\bibfnamefont
  {J.}~\bibnamefont {Hassel}}, \bibinfo {author} {\bibfnamefont
  {S.}~\bibnamefont {Simbierowicz}}, \bibinfo {author} {\bibfnamefont
  {V.}~\bibnamefont {Vesterinen}}, \bibinfo {author} {\bibfnamefont
  {J.}~\bibnamefont {Tuorila}}, \bibinfo {author} {\bibfnamefont
  {T.}~\bibnamefont {Ala-Nissila}}, \ and\ \bibinfo {author} {\bibfnamefont
  {M.}~\bibnamefont {M{\"o}tt{\"o}nen}},\ }\href {\doibase
  10.1038/s41598-018-24449-1} {\bibfield  {journal} {\bibinfo  {journal}
  {Scientific Reports}\ }\textbf {\bibinfo {volume} {8}},\ \bibinfo {pages}
  {6325} (\bibinfo {year} {2018})}\BibitemShut {NoStop}%
\bibitem [{\citenamefont {Rastelli}\ and\ \citenamefont
  {Pop}(2018)}]{rastelliPRB18}%
  \BibitemOpen
  \bibfield  {author} {\bibinfo {author} {\bibfnamefont {G.}~\bibnamefont
  {Rastelli}}\ and\ \bibinfo {author} {\bibfnamefont {I.~M.}\ \bibnamefont
  {Pop}},\ }\href {\doibase 10.1103/PhysRevB.97.205429} {\bibfield  {journal}
  {\bibinfo  {journal} {Phys. Rev. B}\ }\textbf {\bibinfo {volume} {97}},\
  \bibinfo {pages} {205429} (\bibinfo {year} {2018})}\BibitemShut {NoStop}%
\bibitem [{\citenamefont {Carmichael}(2015)}]{carmichaelPRX15}%
  \BibitemOpen
  \bibfield  {author} {\bibinfo {author} {\bibfnamefont {H.~J.}\ \bibnamefont
  {Carmichael}},\ }\href {\doibase 10.1103/PhysRevX.5.031028} {\bibfield
  {journal} {\bibinfo  {journal} {Phys. Rev. X}\ }\textbf {\bibinfo {volume}
  {5}},\ \bibinfo {pages} {031028} (\bibinfo {year} {2015})}\BibitemShut
  {NoStop}%
\bibitem [{\citenamefont {Cao}\ \emph {et~al.}(2016)\citenamefont {Cao},
  \citenamefont {Mahmud},\ and\ \citenamefont {Hafezi}}]{caoPRA16}%
  \BibitemOpen
  \bibfield  {author} {\bibinfo {author} {\bibfnamefont {B.}~\bibnamefont
  {Cao}}, \bibinfo {author} {\bibfnamefont {K.~W.}\ \bibnamefont {Mahmud}}, \
  and\ \bibinfo {author} {\bibfnamefont {M.}~\bibnamefont {Hafezi}},\ }\href
  {\doibase 10.1103/PhysRevA.94.063805} {\bibfield  {journal} {\bibinfo
  {journal} {Phys. Rev. A}\ }\textbf {\bibinfo {volume} {94}},\ \bibinfo
  {pages} {063805} (\bibinfo {year} {2016})}\BibitemShut {NoStop}%
\bibitem [{\citenamefont {Murch}\ \emph {et~al.}(2012)\citenamefont {Murch},
  \citenamefont {Vool}, \citenamefont {Zhou}, \citenamefont {Weber},
  \citenamefont {Girvin},\ and\ \citenamefont {Siddiqi}}]{murchPRL12}%
  \BibitemOpen
  \bibfield  {author} {\bibinfo {author} {\bibfnamefont {K.~W.}\ \bibnamefont
  {Murch}}, \bibinfo {author} {\bibfnamefont {U.}~\bibnamefont {Vool}},
  \bibinfo {author} {\bibfnamefont {D.}~\bibnamefont {Zhou}}, \bibinfo {author}
  {\bibfnamefont {S.~J.}\ \bibnamefont {Weber}}, \bibinfo {author}
  {\bibfnamefont {S.~M.}\ \bibnamefont {Girvin}}, \ and\ \bibinfo {author}
  {\bibfnamefont {I.}~\bibnamefont {Siddiqi}},\ }\href {\doibase
  10.1103/PhysRevLett.109.183602} {\bibfield  {journal} {\bibinfo  {journal}
  {Phys. Rev. Lett.}\ }\textbf {\bibinfo {volume} {109}},\ \bibinfo {pages}
  {183602} (\bibinfo {year} {2012})}\BibitemShut {NoStop}%
\bibitem [{\citenamefont {Miranowicz}\ \emph {et~al.}(2014)\citenamefont
  {Miranowicz}, \citenamefont {Bajer}, \citenamefont {Paprzycka}, \citenamefont
  {Liu}, \citenamefont {Zagoskin},\ and\ \citenamefont
  {Nori}}]{miranowiczPRA14}%
  \BibitemOpen
  \bibfield  {author} {\bibinfo {author} {\bibfnamefont {A.}~\bibnamefont
  {Miranowicz}}, \bibinfo {author} {\bibfnamefont {J.~c.~v.}\ \bibnamefont
  {Bajer}}, \bibinfo {author} {\bibfnamefont {M.}~\bibnamefont {Paprzycka}},
  \bibinfo {author} {\bibfnamefont {Y.-x.}\ \bibnamefont {Liu}}, \bibinfo
  {author} {\bibfnamefont {A.~M.}\ \bibnamefont {Zagoskin}}, \ and\ \bibinfo
  {author} {\bibfnamefont {F.}~\bibnamefont {Nori}},\ }\href {\doibase
  10.1103/PhysRevA.90.033831} {\bibfield  {journal} {\bibinfo  {journal} {Phys.
  Rev. A}\ }\textbf {\bibinfo {volume} {90}},\ \bibinfo {pages} {033831}
  (\bibinfo {year} {2014})}\BibitemShut {NoStop}%
\bibitem [{\citenamefont {Mirrahimi}\ \emph {et~al.}(2014)\citenamefont
  {Mirrahimi}, \citenamefont {Leghtas}, \citenamefont {Albert}, \citenamefont
  {Touzard}, \citenamefont {Schoelkopf}, \citenamefont {Jiang},\ and\
  \citenamefont {Devoret}}]{mirrahimiNJP14}%
  \BibitemOpen
  \bibfield  {author} {\bibinfo {author} {\bibfnamefont {M.}~\bibnamefont
  {Mirrahimi}}, \bibinfo {author} {\bibfnamefont {Z.}~\bibnamefont {Leghtas}},
  \bibinfo {author} {\bibfnamefont {V.~V.}\ \bibnamefont {Albert}}, \bibinfo
  {author} {\bibfnamefont {S.}~\bibnamefont {Touzard}}, \bibinfo {author}
  {\bibfnamefont {R.~J.}\ \bibnamefont {Schoelkopf}}, \bibinfo {author}
  {\bibfnamefont {L.}~\bibnamefont {Jiang}}, \ and\ \bibinfo {author}
  {\bibfnamefont {M.~H.}\ \bibnamefont {Devoret}},\ }\href
  {http://stacks.iop.org/1367-2630/16/i=4/a=045014} {\bibfield  {journal}
  {\bibinfo  {journal} {New Journal of Physics}\ }\textbf {\bibinfo {volume}
  {16}},\ \bibinfo {pages} {045014} (\bibinfo {year} {2014})}\BibitemShut
  {NoStop}%
\bibitem [{\citenamefont {Wang}\ \emph {et~al.}(2016)\citenamefont {Wang},
  \citenamefont {Gao}, \citenamefont {Reinhold}, \citenamefont {Heeres},
  \citenamefont {Ofek}, \citenamefont {Chou}, \citenamefont {Axline},
  \citenamefont {Reagor}, \citenamefont {Blumoff}, \citenamefont {Sliwa},
  \citenamefont {Frunzio}, \citenamefont {Girvin}, \citenamefont {Jiang},
  \citenamefont {Mirrahimi}, \citenamefont {Devoret},\ and\ \citenamefont
  {Schoelkopf}}]{wangSCI16}%
  \BibitemOpen
  \bibfield  {author} {\bibinfo {author} {\bibfnamefont {C.}~\bibnamefont
  {Wang}}, \bibinfo {author} {\bibfnamefont {Y.~Y.}\ \bibnamefont {Gao}},
  \bibinfo {author} {\bibfnamefont {P.}~\bibnamefont {Reinhold}}, \bibinfo
  {author} {\bibfnamefont {R.~W.}\ \bibnamefont {Heeres}}, \bibinfo {author}
  {\bibfnamefont {N.}~\bibnamefont {Ofek}}, \bibinfo {author} {\bibfnamefont
  {K.}~\bibnamefont {Chou}}, \bibinfo {author} {\bibfnamefont {C.}~\bibnamefont
  {Axline}}, \bibinfo {author} {\bibfnamefont {M.}~\bibnamefont {Reagor}},
  \bibinfo {author} {\bibfnamefont {J.}~\bibnamefont {Blumoff}}, \bibinfo
  {author} {\bibfnamefont {K.~M.}\ \bibnamefont {Sliwa}}, \bibinfo {author}
  {\bibfnamefont {L.}~\bibnamefont {Frunzio}}, \bibinfo {author} {\bibfnamefont
  {S.~M.}\ \bibnamefont {Girvin}}, \bibinfo {author} {\bibfnamefont
  {L.}~\bibnamefont {Jiang}}, \bibinfo {author} {\bibfnamefont
  {M.}~\bibnamefont {Mirrahimi}}, \bibinfo {author} {\bibfnamefont {M.~H.}\
  \bibnamefont {Devoret}}, \ and\ \bibinfo {author} {\bibfnamefont {R.~J.}\
  \bibnamefont {Schoelkopf}},\ }\href {\doibase 10.1126/science.aaf2941}
  {\bibfield  {journal} {\bibinfo  {journal} {Science}\ }\textbf {\bibinfo
  {volume} {352}},\ \bibinfo {pages} {1087} (\bibinfo {year}
  {2016})}\BibitemShut {NoStop}%
\bibitem [{\citenamefont {Merzbacher}(1960)}]{merzbacher60}%
  \BibitemOpen
  \bibfield  {author} {\bibinfo {author} {\bibfnamefont {E.}~\bibnamefont
  {Merzbacher}},\ }\href@noop {} {\emph {\bibinfo {title} {Quantum
  Mechanics}}}\ (\bibinfo {year} {1960})\BibitemShut {NoStop}%
\bibitem [{\citenamefont {Carmichael}(2009)}]{carmichael09}%
  \BibitemOpen
  \bibfield  {author} {\bibinfo {author} {\bibfnamefont {H.}~\bibnamefont
  {Carmichael}},\ }\href {https://books.google.com/books?id=sXeMO9EfyUgC}
  {\emph {\bibinfo {title} {Statistical Methods in Quantum Optics 2:
  Non-Classical Fields}}},\ Theoretical and Mathematical Physics\ (\bibinfo
  {publisher} {Springer Berlin Heidelberg},\ \bibinfo {year}
  {2009})\BibitemShut {NoStop}%
\bibitem [{\citenamefont {Miranowicz}\ \emph {et~al.}(2013)\citenamefont
  {Miranowicz}, \citenamefont {Paprzycka}, \citenamefont {Liu}, \citenamefont
  {Bajer},\ and\ \citenamefont {Nori}}]{miranowiczPRA13}%
  \BibitemOpen
  \bibfield  {author} {\bibinfo {author} {\bibfnamefont {A.}~\bibnamefont
  {Miranowicz}}, \bibinfo {author} {\bibfnamefont {M.}~\bibnamefont
  {Paprzycka}}, \bibinfo {author} {\bibfnamefont {Y.-x.}\ \bibnamefont {Liu}},
  \bibinfo {author} {\bibfnamefont {J.~c.~v.}\ \bibnamefont {Bajer}}, \ and\
  \bibinfo {author} {\bibfnamefont {F.}~\bibnamefont {Nori}},\ }\href {\doibase
  10.1103/PhysRevA.87.023809} {\bibfield  {journal} {\bibinfo  {journal} {Phys.
  Rev. A}\ }\textbf {\bibinfo {volume} {87}},\ \bibinfo {pages} {023809}
  (\bibinfo {year} {2013})}\BibitemShut {NoStop}%
\bibitem [{\citenamefont {Koch}\ \emph {et~al.}(2007)\citenamefont {Koch},
  \citenamefont {Yu}, \citenamefont {Gambetta}, \citenamefont {Houck},
  \citenamefont {Schuster}, \citenamefont {Majer}, \citenamefont {Blais},
  \citenamefont {Devoret}, \citenamefont {Girvin},\ and\ \citenamefont
  {Schoelkopf}}]{kochPRA07}%
  \BibitemOpen
  \bibfield  {author} {\bibinfo {author} {\bibfnamefont {J.}~\bibnamefont
  {Koch}}, \bibinfo {author} {\bibfnamefont {T.~M.}\ \bibnamefont {Yu}},
  \bibinfo {author} {\bibfnamefont {J.}~\bibnamefont {Gambetta}}, \bibinfo
  {author} {\bibfnamefont {A.~A.}\ \bibnamefont {Houck}}, \bibinfo {author}
  {\bibfnamefont {D.~I.}\ \bibnamefont {Schuster}}, \bibinfo {author}
  {\bibfnamefont {J.}~\bibnamefont {Majer}}, \bibinfo {author} {\bibfnamefont
  {A.}~\bibnamefont {Blais}}, \bibinfo {author} {\bibfnamefont {M.~H.}\
  \bibnamefont {Devoret}}, \bibinfo {author} {\bibfnamefont {S.~M.}\
  \bibnamefont {Girvin}}, \ and\ \bibinfo {author} {\bibfnamefont {R.~J.}\
  \bibnamefont {Schoelkopf}},\ }\href {\doibase 10.1103/PhysRevA.76.042319}
  {\bibfield  {journal} {\bibinfo  {journal} {Phys. Rev. A}\ }\textbf {\bibinfo
  {volume} {76}},\ \bibinfo {pages} {042319} (\bibinfo {year}
  {2007})}\BibitemShut {NoStop}%
\bibitem [{\citenamefont {Nigg}\ \emph {et~al.}(2012)\citenamefont {Nigg},
  \citenamefont {Paik}, \citenamefont {Vlastakis}, \citenamefont {Kirchmair},
  \citenamefont {Shankar}, \citenamefont {Frunzio}, \citenamefont {Devoret},
  \citenamefont {Schoelkopf},\ and\ \citenamefont {Girvin}}]{niggPRL12}%
  \BibitemOpen
  \bibfield  {author} {\bibinfo {author} {\bibfnamefont {S.~E.}\ \bibnamefont
  {Nigg}}, \bibinfo {author} {\bibfnamefont {H.}~\bibnamefont {Paik}}, \bibinfo
  {author} {\bibfnamefont {B.}~\bibnamefont {Vlastakis}}, \bibinfo {author}
  {\bibfnamefont {G.}~\bibnamefont {Kirchmair}}, \bibinfo {author}
  {\bibfnamefont {S.}~\bibnamefont {Shankar}}, \bibinfo {author} {\bibfnamefont
  {L.}~\bibnamefont {Frunzio}}, \bibinfo {author} {\bibfnamefont {M.~H.}\
  \bibnamefont {Devoret}}, \bibinfo {author} {\bibfnamefont {R.~J.}\
  \bibnamefont {Schoelkopf}}, \ and\ \bibinfo {author} {\bibfnamefont {S.~M.}\
  \bibnamefont {Girvin}},\ }\href {\doibase 10.1103/PhysRevLett.108.240502}
  {\bibfield  {journal} {\bibinfo  {journal} {Phys. Rev. Lett.}\ }\textbf
  {\bibinfo {volume} {108}},\ \bibinfo {pages} {240502} (\bibinfo {year}
  {2012})}\BibitemShut {NoStop}%
\bibitem [{\citenamefont {Kirchmair}\ \emph {et~al.}(2013)\citenamefont
  {Kirchmair}, \citenamefont {Vlastakis}, \citenamefont {Leghtas},
  \citenamefont {Nigg}, \citenamefont {Paik}, \citenamefont {Ginossar},
  \citenamefont {Mirrahimi}, \citenamefont {Frunzio}, \citenamefont {Girvin},\
  and\ \citenamefont {Schoelkopf}}]{kirchmairNAT13}%
  \BibitemOpen
  \bibfield  {author} {\bibinfo {author} {\bibfnamefont {G.}~\bibnamefont
  {Kirchmair}}, \bibinfo {author} {\bibfnamefont {B.}~\bibnamefont
  {Vlastakis}}, \bibinfo {author} {\bibfnamefont {Z.}~\bibnamefont {Leghtas}},
  \bibinfo {author} {\bibfnamefont {S.~E.}\ \bibnamefont {Nigg}}, \bibinfo
  {author} {\bibfnamefont {H.}~\bibnamefont {Paik}}, \bibinfo {author}
  {\bibfnamefont {E.}~\bibnamefont {Ginossar}}, \bibinfo {author}
  {\bibfnamefont {M.}~\bibnamefont {Mirrahimi}}, \bibinfo {author}
  {\bibfnamefont {L.}~\bibnamefont {Frunzio}}, \bibinfo {author} {\bibfnamefont
  {S.~M.}\ \bibnamefont {Girvin}}, \ and\ \bibinfo {author} {\bibfnamefont
  {R.~J.}\ \bibnamefont {Schoelkopf}},\ }\href
  {http://dx.doi.org/10.1038/nature11902} {\bibfield  {journal} {\bibinfo
  {journal} {Nature}\ }\textbf {\bibinfo {volume} {495}},\ \bibinfo {pages}
  {205} (\bibinfo {year} {2013})}\BibitemShut {NoStop}%
\bibitem [{\citenamefont {Boissonneault}\ \emph {et~al.}(2010)\citenamefont
  {Boissonneault}, \citenamefont {Gambetta},\ and\ \citenamefont
  {Blais}}]{boissonneaultPRL10}%
  \BibitemOpen
  \bibfield  {author} {\bibinfo {author} {\bibfnamefont {M.}~\bibnamefont
  {Boissonneault}}, \bibinfo {author} {\bibfnamefont {J.~M.}\ \bibnamefont
  {Gambetta}}, \ and\ \bibinfo {author} {\bibfnamefont {A.}~\bibnamefont
  {Blais}},\ }\href {\doibase 10.1103/PhysRevLett.105.100504} {\bibfield
  {journal} {\bibinfo  {journal} {Phys. Rev. Lett.}\ }\textbf {\bibinfo
  {volume} {105}},\ \bibinfo {pages} {100504} (\bibinfo {year}
  {2010})}\BibitemShut {NoStop}%
\bibitem [{\citenamefont {Bishop}\ \emph {et~al.}(2010)\citenamefont {Bishop},
  \citenamefont {Ginossar},\ and\ \citenamefont {Girvin}}]{bishopPRL10}%
  \BibitemOpen
  \bibfield  {author} {\bibinfo {author} {\bibfnamefont {L.~S.}\ \bibnamefont
  {Bishop}}, \bibinfo {author} {\bibfnamefont {E.}~\bibnamefont {Ginossar}}, \
  and\ \bibinfo {author} {\bibfnamefont {S.~M.}\ \bibnamefont {Girvin}},\
  }\href {\doibase 10.1103/PhysRevLett.105.100505} {\bibfield  {journal}
  {\bibinfo  {journal} {Phys. Rev. Lett.}\ }\textbf {\bibinfo {volume} {105}},\
  \bibinfo {pages} {100505} (\bibinfo {year} {2010})}\BibitemShut {NoStop}%
\bibitem [{\citenamefont {Vool}\ and\ \citenamefont
  {Devoret}(2017)}]{voolIJCT17}%
  \BibitemOpen
  \bibfield  {author} {\bibinfo {author} {\bibfnamefont {U.}~\bibnamefont
  {Vool}}\ and\ \bibinfo {author} {\bibfnamefont {M.}~\bibnamefont {Devoret}},\
  }\href {\doibase 10.1002/cta.2359} {\bibfield  {journal} {\bibinfo  {journal}
  {International Journal of Circuit Theory and Applications}\ }\textbf
  {\bibinfo {volume} {45}},\ \bibinfo {pages} {897} (\bibinfo {year} {2017})},\
  \bibinfo {note} {cta.2359}\BibitemShut {NoStop}%
\bibitem [{\citenamefont {Yurke}\ and\ \citenamefont
  {Denker}(1984)}]{yurkePRA84}%
  \BibitemOpen
  \bibfield  {author} {\bibinfo {author} {\bibfnamefont {B.}~\bibnamefont
  {Yurke}}\ and\ \bibinfo {author} {\bibfnamefont {J.~S.}\ \bibnamefont
  {Denker}},\ }\href {\doibase 10.1103/PhysRevA.29.1419} {\bibfield  {journal}
  {\bibinfo  {journal} {Phys. Rev. A}\ }\textbf {\bibinfo {volume} {29}},\
  \bibinfo {pages} {1419} (\bibinfo {year} {1984})}\BibitemShut {NoStop}%
\bibitem [{\citenamefont {Gardiner}\ and\ \citenamefont
  {Zoller}(2004)}]{gardinerBook04}%
  \BibitemOpen
  \bibfield  {author} {\bibinfo {author} {\bibfnamefont {C.}~\bibnamefont
  {Gardiner}}\ and\ \bibinfo {author} {\bibfnamefont {P.}~\bibnamefont
  {Zoller}},\ }\href@noop {} {\emph {\bibinfo {title} {Quantum Noise: A
  Handbook of Markovian and Non-Markovian Quantum Stochastic Methods with
  Applications to Quantum Optics}}},\ Springer Series in Synergetics\ (\bibinfo
   {publisher} {Springer},\ \bibinfo {address} {New York, NY, USA},\ \bibinfo
  {year} {2004})\BibitemShut {NoStop}%
\bibitem [{\citenamefont {Milburn}\ \emph {et~al.}(2015)\citenamefont
  {Milburn}, \citenamefont {Doppler}, \citenamefont {Holmes}, \citenamefont
  {Portolan}, \citenamefont {Rotter},\ and\ \citenamefont
  {Rabl}}]{milburnPRA15}%
  \BibitemOpen
  \bibfield  {author} {\bibinfo {author} {\bibfnamefont {T.~J.}\ \bibnamefont
  {Milburn}}, \bibinfo {author} {\bibfnamefont {J.}~\bibnamefont {Doppler}},
  \bibinfo {author} {\bibfnamefont {C.~A.}\ \bibnamefont {Holmes}}, \bibinfo
  {author} {\bibfnamefont {S.}~\bibnamefont {Portolan}}, \bibinfo {author}
  {\bibfnamefont {S.}~\bibnamefont {Rotter}}, \ and\ \bibinfo {author}
  {\bibfnamefont {P.}~\bibnamefont {Rabl}},\ }\href {\doibase
  10.1103/PhysRevA.92.052124} {\bibfield  {journal} {\bibinfo  {journal} {Phys.
  Rev. A}\ }\textbf {\bibinfo {volume} {92}},\ \bibinfo {pages} {052124}
  (\bibinfo {year} {2015})}\BibitemShut {NoStop}%
\bibitem [{\citenamefont {Rotter}(2009)}]{rotterJPA09}%
  \BibitemOpen
  \bibfield  {author} {\bibinfo {author} {\bibfnamefont {I.}~\bibnamefont
  {Rotter}},\ }\href {http://stacks.iop.org/1751-8121/42/i=15/a=153001}
  {\bibfield  {journal} {\bibinfo  {journal} {Journal of Physics A:
  Mathematical and Theoretical}\ }\textbf {\bibinfo {volume} {42}},\ \bibinfo
  {pages} {153001} (\bibinfo {year} {2009})}\BibitemShut {NoStop}%
\bibitem [{\citenamefont {Rotter}\ and\ \citenamefont
  {Sadreev}(2005)}]{PhysRevE.71.036227}%
  \BibitemOpen
  \bibfield  {author} {\bibinfo {author} {\bibfnamefont {I.}~\bibnamefont
  {Rotter}}\ and\ \bibinfo {author} {\bibfnamefont {A.~F.}\ \bibnamefont
  {Sadreev}},\ }\href {\doibase 10.1103/PhysRevE.71.036227} {\bibfield
  {journal} {\bibinfo  {journal} {Phys. Rev. E}\ }\textbf {\bibinfo {volume}
  {71}},\ \bibinfo {pages} {036227} (\bibinfo {year} {2005})}\BibitemShut
  {NoStop}%
\bibitem [{Note1()}]{Note1}%
  \BibitemOpen
  \bibinfo {note} {Note that $S_{ij}\propto e^{-\protect \mathaccentV
  {bar}016\gamma t/2}$ has a damping envelope at the average decay rate. In
  particular, when $\gamma _->g$ as $\Delta _b\to 0$ and $g\to 0$, the terms
  $e^{-\protect \mathaccentV {bar}016\gamma t/2}\sinh (\gamma _-t/4)$ and
  $e^{-\protect \mathaccentV {bar}016\gamma t/2}\cosh (\gamma _-t/4)$ remain
  finite because $\protect \mathaccentV {bar}016\gamma >\gamma
  _-/2$.}\BibitemShut {Stop}%
\bibitem [{Note2()}]{Note2}%
  \BibitemOpen
  \bibinfo {note} {Note that, in contrast to normal modes, these dissipative
  modes are not orthorgonal because $\vb U$ is not a unitary
  matrix.}\BibitemShut {Stop}%
\bibitem [{\citenamefont {Berry}(2004)}]{berryCJP2004}%
  \BibitemOpen
  \bibfield  {author} {\bibinfo {author} {\bibfnamefont {M.}~\bibnamefont
  {Berry}},\ }\href {\doibase 10.1023/B:CJOP.0000044002.05657.04} {\bibfield
  {journal} {\bibinfo  {journal} {Czechoslovak Journal of Physics}\ }\textbf
  {\bibinfo {volume} {54}},\ \bibinfo {pages} {1039} (\bibinfo {year}
  {2004})}\BibitemShut {NoStop}%
\bibitem [{\citenamefont {Dembowski}\ \emph {et~al.}(2001)\citenamefont
  {Dembowski}, \citenamefont {Gr\"af}, \citenamefont {Harney}, \citenamefont
  {Heine}, \citenamefont {Heiss}, \citenamefont {Rehfeld},\ and\ \citenamefont
  {Richter}}]{dembowskiPRL01}%
  \BibitemOpen
  \bibfield  {author} {\bibinfo {author} {\bibfnamefont {C.}~\bibnamefont
  {Dembowski}}, \bibinfo {author} {\bibfnamefont {H.-D.}\ \bibnamefont
  {Gr\"af}}, \bibinfo {author} {\bibfnamefont {H.~L.}\ \bibnamefont {Harney}},
  \bibinfo {author} {\bibfnamefont {A.}~\bibnamefont {Heine}}, \bibinfo
  {author} {\bibfnamefont {W.~D.}\ \bibnamefont {Heiss}}, \bibinfo {author}
  {\bibfnamefont {H.}~\bibnamefont {Rehfeld}}, \ and\ \bibinfo {author}
  {\bibfnamefont {A.}~\bibnamefont {Richter}},\ }\href {\doibase
  10.1103/PhysRevLett.86.787} {\bibfield  {journal} {\bibinfo  {journal} {Phys.
  Rev. Lett.}\ }\textbf {\bibinfo {volume} {86}},\ \bibinfo {pages} {787}
  (\bibinfo {year} {2001})}\BibitemShut {NoStop}%
\bibitem [{\citenamefont {Uzdin}\ \emph {et~al.}(2011)\citenamefont {Uzdin},
  \citenamefont {Mailybaev},\ and\ \citenamefont {Moiseyev}}]{uzdin11}%
  \BibitemOpen
  \bibfield  {author} {\bibinfo {author} {\bibfnamefont {R.}~\bibnamefont
  {Uzdin}}, \bibinfo {author} {\bibfnamefont {A.}~\bibnamefont {Mailybaev}}, \
  and\ \bibinfo {author} {\bibfnamefont {N.}~\bibnamefont {Moiseyev}},\ }\href
  {http://stacks.iop.org/1751-8121/44/i=43/a=435302} {\bibfield  {journal}
  {\bibinfo  {journal} {Journal of Physics A: Mathematical and Theoretical}\
  }\textbf {\bibinfo {volume} {44}},\ \bibinfo {pages} {435302} (\bibinfo
  {year} {2011})}\BibitemShut {NoStop}%
\bibitem [{\citenamefont {{Clerk}}\ \emph {et~al.}(2010)\citenamefont
  {{Clerk}}, \citenamefont {{Devoret}}, \citenamefont {{Girvin}}, \citenamefont
  {{Marquardt}},\ and\ \citenamefont {{Schoelkopf}}}]{clerkRMP10}%
  \BibitemOpen
  \bibfield  {author} {\bibinfo {author} {\bibfnamefont {A.~A.}\ \bibnamefont
  {{Clerk}}}, \bibinfo {author} {\bibfnamefont {M.~H.}\ \bibnamefont
  {{Devoret}}}, \bibinfo {author} {\bibfnamefont {S.~M.}\ \bibnamefont
  {{Girvin}}}, \bibinfo {author} {\bibfnamefont {F.}~\bibnamefont
  {{Marquardt}}}, \ and\ \bibinfo {author} {\bibfnamefont {R.~J.}\ \bibnamefont
  {{Schoelkopf}}},\ }\href {\doibase 10.1103/RevModPhys.82.1155} {\bibfield
  {journal} {\bibinfo  {journal} {Reviews of Modern Physics}\ }\textbf
  {\bibinfo {volume} {82}},\ \bibinfo {pages} {1155} (\bibinfo {year}
  {2010})}\BibitemShut {NoStop}%
\bibitem [{\citenamefont {Walls}\ and\ \citenamefont
  {Milburn}(2012)}]{wallsBook2012}%
  \BibitemOpen
  \bibfield  {author} {\bibinfo {author} {\bibfnamefont {D.}~\bibnamefont
  {Walls}}\ and\ \bibinfo {author} {\bibfnamefont {G.}~\bibnamefont
  {Milburn}},\ }\href {https://books.google.com/books?id=o6nrCAAAQBAJ} {\emph
  {\bibinfo {title} {Quantum Optics}}},\ Springer Study Edition\ (\bibinfo
  {publisher} {Springer Berlin Heidelberg},\ \bibinfo {year}
  {2012})\BibitemShut {NoStop}%
\bibitem [{\citenamefont {Eilbeck}\ \emph {et~al.}(1985)\citenamefont
  {Eilbeck}, \citenamefont {Lomdahl},\ and\ \citenamefont
  {Scott}}]{eilbeckPD85}%
  \BibitemOpen
  \bibfield  {author} {\bibinfo {author} {\bibfnamefont {J.}~\bibnamefont
  {Eilbeck}}, \bibinfo {author} {\bibfnamefont {P.}~\bibnamefont {Lomdahl}}, \
  and\ \bibinfo {author} {\bibfnamefont {A.}~\bibnamefont {Scott}},\ }\href
  {\doibase http://dx.doi.org/10.1016/0167-2789(85)90012-0} {\bibfield
  {journal} {\bibinfo  {journal} {Physica D: Nonlinear Phenomena}\ }\textbf
  {\bibinfo {volume} {16}},\ \bibinfo {pages} {318 } (\bibinfo {year}
  {1985})}\BibitemShut {NoStop}%
\bibitem [{Note3()}]{Note3}%
  \BibitemOpen
  \bibinfo {note} {Note that even with $\sim 10$ initial photons, the basis
  required for this simulation has $\sim 10^2$ states, and Eq.~\protect \textup
  {\hbox {\mathsurround \z@ \protect \normalfont (\ignorespaces \ref
  {master}\unskip \@@italiccorr )}} involves $\sim 10^4$ coupled differential
  equations.}\BibitemShut {Stop}%
\bibitem [{\citenamefont {Johansson}\ \emph {et~al.}(2012)\citenamefont
  {Johansson}, \citenamefont {Nation},\ and\ \citenamefont
  {Nori}}]{johanssonCPC12}%
  \BibitemOpen
  \bibfield  {author} {\bibinfo {author} {\bibfnamefont {J.~R.}\ \bibnamefont
  {Johansson}}, \bibinfo {author} {\bibfnamefont {P.~D.}\ \bibnamefont
  {Nation}}, \ and\ \bibinfo {author} {\bibfnamefont {F.}~\bibnamefont
  {Nori}},\ }\href {\doibase http://dx.doi.org/10.1016/j.cpc.2012.02.021}
  {\bibfield  {journal} {\bibinfo  {journal} {Computer Physics Communications}\
  }\textbf {\bibinfo {volume} {183}},\ \bibinfo {pages} {1760} (\bibinfo {year}
  {2012})}\BibitemShut {NoStop}%
\bibitem [{\citenamefont {Tuorila}\ \emph {et~al.}(2017)\citenamefont
  {Tuorila}, \citenamefont {Partanen}, \citenamefont {Ala-Nissila},\ and\
  \citenamefont {M{\"o}tt{\"o}nen}}]{tuorilaNPJ17}%
  \BibitemOpen
  \bibfield  {author} {\bibinfo {author} {\bibfnamefont {J.}~\bibnamefont
  {Tuorila}}, \bibinfo {author} {\bibfnamefont {M.}~\bibnamefont {Partanen}},
  \bibinfo {author} {\bibfnamefont {T.}~\bibnamefont {Ala-Nissila}}, \ and\
  \bibinfo {author} {\bibfnamefont {M.}~\bibnamefont {M{\"o}tt{\"o}nen}},\
  }\href {\doibase 10.1038/s41534-017-0027-1} {\bibfield  {journal} {\bibinfo
  {journal} {npj Quantum Information}\ }\textbf {\bibinfo {volume} {3}},\
  \bibinfo {pages} {27} (\bibinfo {year} {2017})}\BibitemShut {NoStop}%
\bibitem [{\citenamefont {Blais}\ \emph {et~al.}(2007)\citenamefont {Blais},
  \citenamefont {Gambetta}, \citenamefont {Wallraff}, \citenamefont {Schuster},
  \citenamefont {Girvin}, \citenamefont {Devoret},\ and\ \citenamefont
  {Schoelkopf}}]{blaisPRA07}%
  \BibitemOpen
  \bibfield  {author} {\bibinfo {author} {\bibfnamefont {A.}~\bibnamefont
  {Blais}}, \bibinfo {author} {\bibfnamefont {J.}~\bibnamefont {Gambetta}},
  \bibinfo {author} {\bibfnamefont {A.}~\bibnamefont {Wallraff}}, \bibinfo
  {author} {\bibfnamefont {D.~I.}\ \bibnamefont {Schuster}}, \bibinfo {author}
  {\bibfnamefont {S.~M.}\ \bibnamefont {Girvin}}, \bibinfo {author}
  {\bibfnamefont {M.~H.}\ \bibnamefont {Devoret}}, \ and\ \bibinfo {author}
  {\bibfnamefont {R.~J.}\ \bibnamefont {Schoelkopf}},\ }\href {\doibase
  10.1103/PhysRevA.75.032329} {\bibfield  {journal} {\bibinfo  {journal} {Phys.
  Rev. A}\ }\textbf {\bibinfo {volume} {75}},\ \bibinfo {pages} {032329}
  (\bibinfo {year} {2007})}\BibitemShut {NoStop}%
\bibitem [{\citenamefont {Wallraff}\ \emph {et~al.}(2007)\citenamefont
  {Wallraff}, \citenamefont {Schuster}, \citenamefont {Blais}, \citenamefont
  {Gambetta}, \citenamefont {Schreier}, \citenamefont {Frunzio}, \citenamefont
  {Devoret}, \citenamefont {Girvin},\ and\ \citenamefont
  {Schoelkopf}}]{wallraffPRL07}%
  \BibitemOpen
  \bibfield  {author} {\bibinfo {author} {\bibfnamefont {A.}~\bibnamefont
  {Wallraff}}, \bibinfo {author} {\bibfnamefont {D.~I.}\ \bibnamefont
  {Schuster}}, \bibinfo {author} {\bibfnamefont {A.}~\bibnamefont {Blais}},
  \bibinfo {author} {\bibfnamefont {J.~M.}\ \bibnamefont {Gambetta}}, \bibinfo
  {author} {\bibfnamefont {J.}~\bibnamefont {Schreier}}, \bibinfo {author}
  {\bibfnamefont {L.}~\bibnamefont {Frunzio}}, \bibinfo {author} {\bibfnamefont
  {M.~H.}\ \bibnamefont {Devoret}}, \bibinfo {author} {\bibfnamefont {S.~M.}\
  \bibnamefont {Girvin}}, \ and\ \bibinfo {author} {\bibfnamefont {R.~J.}\
  \bibnamefont {Schoelkopf}},\ }\href {\doibase 10.1103/PhysRevLett.99.050501}
  {\bibfield  {journal} {\bibinfo  {journal} {Phys. Rev. Lett.}\ }\textbf
  {\bibinfo {volume} {99}},\ \bibinfo {pages} {050501} (\bibinfo {year}
  {2007})}\BibitemShut {NoStop}%
\bibitem [{\citenamefont {Magnard}\ \emph {et~al.}(2018)\citenamefont
  {Magnard}, \citenamefont {Kurpiers}, \citenamefont {Royer}, \citenamefont
  {Walter}, \citenamefont {Besse}, \citenamefont {Gasparinetti}, \citenamefont
  {Pechal}, \citenamefont {Heinsoo}, \citenamefont {Storz}, \citenamefont
  {Blais},\ and\ \citenamefont {Wallraff}}]{magnardPRL18}%
  \BibitemOpen
  \bibfield  {author} {\bibinfo {author} {\bibfnamefont {P.}~\bibnamefont
  {Magnard}}, \bibinfo {author} {\bibfnamefont {P.}~\bibnamefont {Kurpiers}},
  \bibinfo {author} {\bibfnamefont {B.}~\bibnamefont {Royer}}, \bibinfo
  {author} {\bibfnamefont {T.}~\bibnamefont {Walter}}, \bibinfo {author}
  {\bibfnamefont {J.-C.}\ \bibnamefont {Besse}}, \bibinfo {author}
  {\bibfnamefont {S.}~\bibnamefont {Gasparinetti}}, \bibinfo {author}
  {\bibfnamefont {M.}~\bibnamefont {Pechal}}, \bibinfo {author} {\bibfnamefont
  {J.}~\bibnamefont {Heinsoo}}, \bibinfo {author} {\bibfnamefont
  {S.}~\bibnamefont {Storz}}, \bibinfo {author} {\bibfnamefont
  {A.}~\bibnamefont {Blais}}, \ and\ \bibinfo {author} {\bibfnamefont
  {A.}~\bibnamefont {Wallraff}},\ }\href {\doibase
  10.1103/PhysRevLett.121.060502} {\bibfield  {journal} {\bibinfo  {journal}
  {Phys. Rev. Lett.}\ }\textbf {\bibinfo {volume} {121}},\ \bibinfo {pages}
  {060502} (\bibinfo {year} {2018})}\BibitemShut {NoStop}%
\end{thebibliography}%
\end{document}